\documentclass[aps,prl,twocolumn,superscriptaddress,showpacs,floatfix,longbibliography]{revtex4-2}
\usepackage{graphicx} 
\usepackage{amsmath,amssymb,amsfonts,float,graphics,epsfig,epstopdf,color,verbatim,tabularx,bm,multirow,appendix}
\usepackage{tikz}
\usetikzlibrary{arrows.meta,positioning}
\usepackage{amsmath,amssymb,graphicx}
\usepackage{wasysym}
\usepackage[utf8]{inputenc}
\usepackage[T1]{fontenc}
\usepackage{xcolor}
\usepackage{dsfont}
\usepackage{textcomp}
\usepackage{bm}

\usepackage{graphicx}%
\usepackage{xcolor}
\usepackage{dcolumn}
\usepackage{bm}
\usepackage{xspace}
\usepackage{time}
\usepackage{booktabs}
\usepackage{multirow}
\usepackage{makecell}
\usepackage[normalem]{ulem} 

\usepackage{comment}

\pdfoutput=1
\usepackage{color}
\definecolor{LinkColor}{rgb}{0.256,0.439,0.588}
\usepackage{hyperref}
\hypersetup{
colorlinks=true,
citecolor=LinkColor,
linkcolor=LinkColor,
urlcolor=LinkColor
}
\usepackage{cleveref}
\Crefname{equation}{Eq.}{Eqs.}
\Crefname{figure}{Fig.}{Figs.}

\newcommand{\newsect}[1]{\noindent \textit{\textcolor{blue}{#1.--}}}

\newcommand{\av}[1]{\langle#1\rangle}

\newcommand{\new}[1]{{#1}} 

\graphicspath{{./figures/}{./trees/}}

\begin{document}
\title{Many-Anyon Braiding in Non-Abelian Fractional Quantum Hall Effect with Hybrid Monte Carlo Simulation}

\author{Ting-Tung Wang}
\thanks{These authors contributed equally.}
\affiliation{Department of Physics and HK Institute of Quantum Science \& Technology, The University of Hong Kong, Pokfulam Road,  Hong Kong SAR, China}
\affiliation{State Key Laboratory of Optical Quantum Materials, The University of Hong Kong, Pokfulam Road,  Hong Kong SAR, China}

\author{Ha Quang Trung}
\thanks{These authors contributed equally.}
\affiliation{Division of Physics and Applied Physics, Nanyang Technological University, Singapore 637371, Singapore}

\author{Qianhui Xu}
\affiliation{Division of Physics and Applied Physics, Nanyang Technological University, Singapore 637371, Singapore}

\author{Min Long}
\affiliation{Department of Physics and HK Institute of Quantum Science \& Technology, The University of Hong Kong, Pokfulam Road,  Hong Kong SAR, China}
\affiliation{State Key Laboratory of Optical Quantum Materials, The University of Hong Kong, Pokfulam Road,  Hong Kong SAR, China}

\author{Bo Yang}
\email{yang.bo@ntu.edu.sg}
\affiliation{Division of Physics and Applied Physics, Nanyang Technological University, Singapore 637371, Singapore}

\author{Zi Yang Meng}
\email{zymeng@hku.hk}
\affiliation{Department of Physics and HK Institute of Quantum Science \& Technology, The University of Hong Kong, Pokfulam Road,  Hong Kong SAR, China}
\affiliation{State Key Laboratory of Optical Quantum Materials, The University of Hong Kong, Pokfulam Road,  Hong Kong SAR, China}

\date{\today}

\begin{abstract}
We use the hybrid Monte Carlo method to efficiently compute the many-anyon non-Abelian braiding matrices for different Moore-Read quasiholes braiding schemes. \new{Using both a traditional two-anyon braiding scheme and a newly proposed braiding scheme based on global rotation, we approximate the non-Abelian characteristic length of Moore-Read quasiholes around $2.8$ times the magnetic length.} 
We also apply this method to braiding schemes involving six anyons and explicitly show that the well-known ``even-odd effect'' can be flipped by tuning electrostatic potentials.
This phenomenon emphasizes the importance of accounting for all anyons within a non-Abelian system, even if the braiding involves only several anyons locally. Our results are relevant to the experimental realization of non-Abelian physics in fractional quantum Hall and other analogous systems, including the fast-growing field of fractional quantum anomalous Hall states in moir\'e materials.

\end{abstract}

\maketitle

\newsect{Introduction and Main Results} The search for emergent non-Abelian anyons in condensed matter systems has been actively pursued since their theoretical proposal~\cite{moore1991nonabelions,wen1991non}. This pursuit is strongly motivated by the potential for fault-tolerant quantum computation: quantum information encoded in the degenerate fusion space of multiple non-Abelian anyons can be manipulated through braiding operations, which depend only on the topology of the braiding trajectories and are therefore intrinsically robust against local perturbations~\cite{Kitaev1997Fault,preskill1999lecture,Nayak2008,sarma2015majorana,aghaee2023inas,youvan2025microsoft,aasen2025roadmap}. This intriguing property is a direct result of the configuration space of identical particles forming a representation of the braid group, giving rise to mutual statistics in two dimensions that are much richer than their three-dimensional counterparts (namely bosons and fermions)~\cite{Leinaas1977}.

Among the quantum systems capable of hosting anyons, the fractional quantum Hall (FQH) systems are arguably the most extensively studied and the best established experimentally~\cite{tsui1982two,ma2022fractional,cage2012quantum,Feldman2021}. More recently, fractional quantum anomalous Hall (FQAH) states have been observed in two-dimensional moir\'e
materials without an external magnetic field~\cite{Cai2023_signature_fqah_mote2,Park2023_observation_fqah_mote2,Zeng2023_thermo_evidence_fqah_mote2,Xu2023_Observation_FQAH_tMote2,Lu2024_FQAH_multilayer_graphene,luExtended2025,park2025observation}, providing zero-field analogues of odd-denominator Abelian FQH phases. Even-denominator states have also been reported in bilayer-graphene-based van der Waals heterostructures, together with experimental signatures compatible with non-Abelian anyons~\cite{kim2026aharonov}. Superconductivity potentially associated with the pairing of anyonic excitations has likewise been reported in related systems~\cite{xuSignatures2025}. These developments substantially broaden the range of material platforms in which anyons may be created, detected, and manipulated. 

Theoretically, non-Abelian FQH model states include the Moore-Read (MR, or ``Pfaffian'') state~\cite{pan1999exact,morf2002excitation,ma2022fractional,morf1998transition,pakrouski2015phase,peterson2008finite} and other states in the Read-Rezayi series, including the ``Fibonacci'' states whose anyons are capable of \emph{universal} quantum computation~\cite{read1999beyond,Read1996}. Among the proposed non-Abelian FQH states, the MR state is currently the most promising for experimental realization: together with its particle-hole conjugate (the ``anti-Pfaffian'') and the particle-hole-symmetric Pfaffian, it is a leading candidate for the even-denominator plateau at $\nu=5/2$ observed in experiments~\cite{willett1987observation,ma2022fractional,Manna2018}. \new{Thermal Hall measurements have provided initial evidence for non-Abelian neutral modes, but several open issues remain, such as controlling the effects of thermalization and resolving discrepancies between theory and experiment~\cite{paul2024topological,banerjee2018observation,ma2019partial,boivin2026bulk}. Moreover,} although several experimental protocols have been proposed to detect signatures of non-Abelian braiding in the Moore-Read (MR) state, including interferometry and anyon collider measurements~\cite{bonderson2006detecting,bonderson2008interferometry,Lee2022}, \new{a direct observation of non-Abelian statistics has yet to be reported.}

The challenges of experimentally realizing non-Abelian braiding motivate us to study many-anyon braiding features, especially their microscopic mechanism, beyond the ideal topological description. Braiding matrices are conventionally obtained from conformal field theory (CFT)~\cite{moore1989classical,moore1991nonabelions,read1999beyond,Read1996,ardonne2007wavefunctions} or modular tensor categories~\cite{moore1991nonabelions,Read1992,read1999beyond,nayak19962n}, which characterize the universal braiding properties in the limit of well-separated, point-like anyons. Real FQH quasiholes, however, have a finite spatial extent and cannot in general be kept arbitrarily far apart. Their finite size gives rise to nonuniversal corrections to anyon dynamics and statistics~\cite{johri2014quasiholes,Comparin2021,trung2023spin,iyer2024finite,trung2025long,xu2025dynamics,gattu2025molecular}, and also determines how they respond to realistic electron-electron interactions, impurities, and the local trapping potentials required to manipulate and braid them. Under these conditions, quasiholes can form bound states~\cite{Parameswaran_2011_Weakly,Parameswaran_2012_Typology}, and preferentially fuse into different channels depending on the microscopic energetics~\cite{xu2025dynamics}. It is therefore important to determine how such interactions, trapping potentials, and fusion processes modify the braiding matrices beyond their universal topological limit, which requires large-scale microscopic numerical calculations.

The four-quasihole braiding in the MR state has been studied in previous simulations~\cite{Tserkovnyak2003,Manna_2019_Quasielectrons,wang2026hybrid}, but microscopic calculations for six or more quasiholes remain unexplored. In the ideal limit of well-separated quasiholes, a braid involving four quasiholes is commonly treated as a local process, independent of the remaining quasiholes, which are ignored by taking a partial trace~\cite{bonderson2006detecting,bonderson2009splitting}. 
Nevertheless, fusion processes in the background can influence a local braid even at large separation, requiring all quasiholes to be retained in a microscopic treatment~\cite{trung2025long,hu2025high}. Many-anyon protocols may also be experimentally useful because several quasiholes could be manipulated in a single operation, potentially simplifying the realization of single-qubit gates, although the total number of quasiholes may not be precisely controlled. 

To overcome the numerical challenges arising from simulating the MR state, which involves the computation of the Pfaffian, we employ the hybrid Monte Carlo (HMC) algorithm that introduces auxiliary momenta and uses Hamiltonian dynamics to generate global, gradient-guided updates of all electron coordinates~\cite{wang2026hybrid}. Compared with conventional local Monte Carlo sampling, these collective updates substantially reduce autocorrelation and enable reliable evaluation of physical observables at much larger system sizes~\cite{fengScalable2025,huangAngle2025,liaoNumerical2026,lunts2023non,patel2024strange,Ostmeyer_2020_uov,Ostmeyer_2021_efs,Ostmeyer_2024_dhw,Krieg_2018_pqh,shanTrion2026,ShanMomentum2026}. Although the model wavefunctions are not exact eigenstates of realistic experimental Hamiltonians, projecting the Coulomb interaction onto the model quasihole manifold captures its leading-order effect within this low-energy subspace~\cite{Wan2008,Tserkovnyak2003,Prodan_Mapping_2009}.

In this Letter, we compute the non-Abelian braiding matrices of MR systems with four and six quasiholes, using HMC applied to the first-quantized model wavefunctions. Beyond verifying theoretical predictions at much larger system sizes than previously reported, these high-precision numerical computation unambiguously demonstrates that quasihole braiding can qualitatively depend on the fusion dynamics of distant $\sigma$-quasiholes. This dependence reflects the long-range entanglement of non-Abelian topological order and implies that reliable anyon manipulation requires control of the global fusion configuration. Access to larger system sizes further enables us to systematically investigate how this fusion dependence is controlled by competition between the Coulomb interaction and a one-body trapping potential. We also show numerically that braiding by a global rotation of multiple quasiholes strongly suppresses the nonuniversal corrections due to finite anyon separation, \new{and we extract the length scale on which these corrections decay}.

\begin{figure}[htp!]
    \centering
\includegraphics[width=\columnwidth]{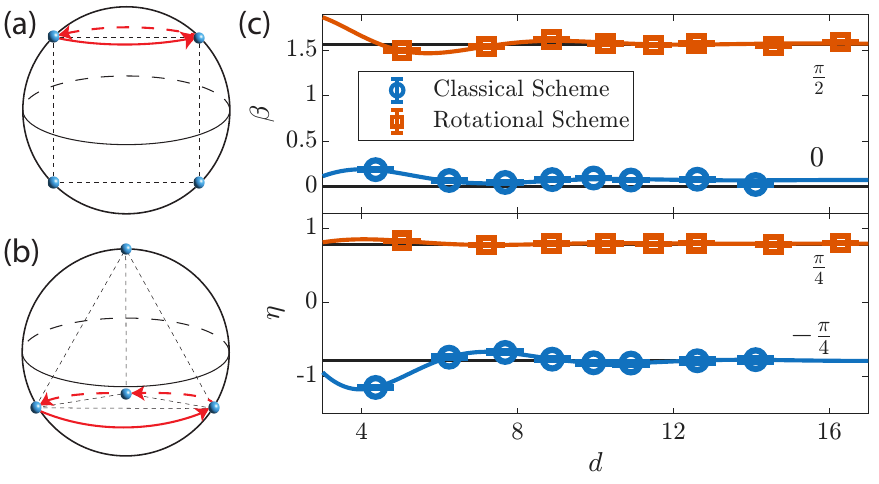}
    \caption{\textbf{Improved convergence of the rotational braiding scheme.} Schematics of (a) the classical and (b) the rotational braiding schemes. The blue dots mark the quasihole positions, while the red arrows indicate their braiding trajectories. (c) Angles $\beta$ and $\eta$ of \Cref{eq:braiding_parameterization} versus the chord distance $d$ (in units of $\ell_B$) for the classical (blue circles) and rotational (orange squares) schemes. Horizontal lines: CFT predictions; curves: joint fit to \Cref{eq:damped oscillation}.}
    \label{fig:schemes_compare}
\end{figure}

\newsect{Robustness of rotational braiding scheme} We first examine finite-size effects in the non-Abelian braiding matrix. For Abelian quasiholes, such effects are relatively well understood: non-topological corrections to the Abelian braiding phase generally arise from two sources, finite separation~\cite{arovas1984fractional,Comparin2021} or shape deformation of the individual anyons~\cite{trung2023spin}. The latter is present even if the anyons are infinitely far apart. In contrast, finite-size corrections to the non-Abelian part of the braiding have been characterized much less systematically~\cite{Tserkovnyak2003,baraban2009numerical,Bonderson2011,wu2014braiding}, as this requires accurate calculations at large system sizes, which the HMC method makes possible~\cite{wang2026hybrid}.

We consider the model wavefunction for the MR state with two additional flux quanta, corresponding to four quasiholes of charge $e/4$~\cite{moore1991nonabelions,Bonderson2011}. Each quasihole is an Ising anyon $\sigma$ with fusion rule $\sigma\times\sigma=1+\psi$, where $1$ is the vacuum and $\psi$ a neutral fermion. As shown in Fig.~\ref{fig:schemes_compare}, the quasiholes are initially placed at four distinct positions on the sphere~\cite{haldane1983fractional} and then moved along trajectories that return the quasihole configuration to itself up to a permutation of the quasiholes. Among the many possible such braiding schemes, we focus on two representative ones: (i) Place the four quasiholes at the vertices of a square inscribed in a great circle of the sphere, and exchange two neighboring quasiholes along a circle of latitude while keeping the other two fixed (see \Cref{fig:schemes_compare}(a)). We refer to this as the ``classical braiding scheme''. (ii) Place the four quasiholes at the vertices of a regular tetrahedron inscribed in the sphere, with one at the north pole, and rotate the entire system by $2\pi/3$ about the $z$-axis (see \Cref{fig:schemes_compare}(b)). This cyclically permutes the three lower quasiholes while the fourth remains at the north pole. We refer to this as the ``rotational braiding scheme''.

The classical braiding scheme is commonly studied in the literature. It is the simplest braiding scheme, exchanging only one pair, and the braiding outcome for MR quasiholes is well known~\cite{Tserkovnyak2003,read1999beyond,Bonderson2011,bonderson2006detecting,Manna_2019_Quasielectrons}. In contrast, the rotational braiding scheme is more complex, composed of four elementary exchanges, but it offers several advantages. First, the rotational symmetry of the system substantially reduces the computational cost of evaluating the Berry matrix: in the discretized braiding path, each segment of the rotational path is related to the others by symmetry, so the overlap calculations need to be performed only for a single representative step and then reused along the entire braid~\cite{wang2026hybrid,seesup}. Second, as we shall show here, this braiding scheme is much more robust against finite anyon separation.

\new{
To quantify the finite-size corrections, we parametrize the computed braiding matrix as
\begin{equation}
    U
    =
    e^{i\chi}
    \begin{pmatrix}
        e^{i\eta}\cos\left(\dfrac{\beta}{2}\right)
        &
        i e^{-i\epsilon/2}
        \sin\left(\dfrac{\beta}{2}\right)
        \\[4pt]
        i e^{i\epsilon/2}
        \sin\left(\dfrac{\beta}{2}\right)
        &
        e^{-i\eta}\cos\left(\dfrac{\beta}{2}\right)
    \end{pmatrix},
    \label{eq:braiding_parameterization}
\end{equation}
in the basis labeled by the fusion channel $a\in\{1,\psi\}$ of a pair of quasiholes. Here, $2\eta$ is the relative phase between the two diagonal elements, $\beta$ controls the mixing between the two fusion channels, $\epsilon$ is the relative phase of the off-diagonal elements, and $\chi$ is an overall phase that includes the nonuniversal Aharonov-Bohm and spin-transport contributions.
Once the two basis states are fixed to the $1$ and $\psi$ fusion channels, $\eta$ and $\beta$ are uniquely determined, whereas $\epsilon$ depends on the phase convention of the basis states. The CFT predicts $(\eta,\beta)=(-\pi/4,0)$ for the classical scheme and $(\pi/4,\pi/2)$ for the rotational scheme~\cite{seesup}.

Figure~\ref{fig:schemes_compare}(c) shows $\eta$ and $\beta$ as functions of the chord distance $d$ between neighboring quasiholes for $N=10$ to $100$ electrons. In both schemes, the two angles approach their CFT values through deviations that oscillate and decay with $d$. This behavior is well described by an exponentially damped oscillation~\cite{ji2024universal}:
\begin{equation}
\label{eq:damped oscillation}
f(d)=f_\infty+A e^{-(d-d_0)/\xi}\cos(kd-\vartheta),
\end{equation} 
where $f$ stands for $\eta$ or $\beta$, $f_\infty$ is its asymptotic value, $A$ is the amplitude at the smallest separation $d_0$, and $\vartheta$ is a phase. A joint fit of all four data sets in \Cref{fig:schemes_compare}(c) with a common decay length and wave number yields $\xi=2.77(6)\,\ell_B$ and $k=0.96(2)\,\ell_B^{-1}$, where $\ell_B$ is the magnetic length~\cite{seesup}. This $\xi$ agrees with the decay length $\approx2.7\,\ell_B$ of the two-quasihole (Abelian) braiding phase~\cite{baraban2009numerical} and with the MR quasihole size, $\approx2.6$--$2.8\,\ell_B$, obtained from matrix-product states~\cite{wu2014braiding}. To our knowledge, this is the first direct demonstration that finite-size corrections to the non-Abelian braiding matrix decay on the same length scale as their Abelian counterpart.

The two schemes nevertheless differ sharply in deviation from theoretical value. The fitting results of parameter $A$ in \Cref{eq:damped oscillation} for $\eta$ show the amplitude of oscillation in the rotational braiding scheme is one order of magnitude smaller compared to in the classical scheme (0.044 compared to 0.362). 
The improved convergence is particularly notable because the rotational protocol corresponds to a more complicated braid composed of four elementary exchanges (see the tree diagram in SM~\cite{seesup}). We attribute it to the symmetry of the tetrahedral configuration and the collective rotation: the relative distance between any pair of quasiholes is preserved at every point along the path, so the finite-size distortion of the quasihole states is identical, up to a rigid rotation, at every step of the braid and does not generate segment-dependent corrections to the Berry connection~\cite{seesup}. In the classical scheme, in contrast, the separations between the moving and the fixed quasiholes vary along the path. The rotational scheme therefore provides a substantially more reliable route to extracting the non-Abelian braiding matrix from finite-size calculations, while simultaneously reducing the numerical cost by leveraging rotational symmetry.
}

\newsect{Fusion channel selection by electrostatic potential}
Having established the advantages of the rotational braiding scheme, we now use global-rotation braids to study how fusion affects braiding in systems of four and six quasiholes. We focus on the problem discussed in Ref.~\cite{trung2025long}, namely how quasihole dynamics and trapping potentials modify non-Abelian braiding.
Our analysis is carried out within the MR quasihole manifold. For $2n$ quasiholes on the sphere, a conformal-block wavefunction can be written as
\begin{widetext}
\begin{equation}
\label{eq:MR quasihole}
\Psi_{\mathrm{MR}}(\{z_i\},\{\eta_k\})
=\mathrm{Pf}\!\left(\frac{\prod_{k=1}^n(z_i-\eta_k)(z_j-\eta_{k+n})+(i\leftrightarrow j)}{z_i-z_j}\right)
\prod_{i<j}(z_i-z_j)^2
\prod_i\left(\frac{1}{1+|z_i|^2}\right)^S ,
\end{equation}
\end{widetext}
\new{Here, $z_i$ and $\eta_k$ are the stereographic coordinates of the $i$th electron and the $k$th quasihole, respectively, $N$ is the number of electrons, and $2S=2N+n-3$ is the number of magnetic flux quanta threading the sphere. We set $\ell_B=1$ hereafter.} Two quasiholes brought to the same position can fuse into either the vacuum channel $1$ or the fermion channel $\psi$. Although the two outcomes carry the same total charge $e/2$, they have distinct local electron-density and energy profiles.

\begin{figure}
    \centering
    \includegraphics[width=\linewidth]{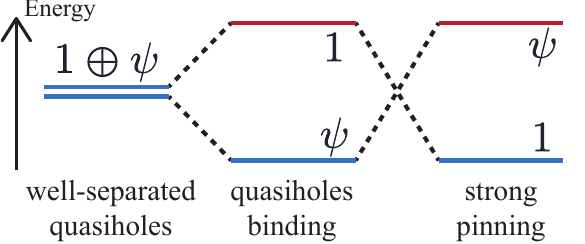}
    \caption{\textbf{Schematic of fusion-channel selection.} For a Moore-Read state with $2n$ well-separated quasiholes, the fusion manifold is $2^{n-1}$-fold degenerate. Bringing two quasiholes together lifts this degeneracy and splits the manifold into two $2^{n-2}$-fold sectors corresponding to the fusion outcomes $1$ and $\psi$. The Coulomb interaction favors the $\psi$ sector, placing it at lower energy, labeled by the blue level in the middle panel. A sufficiently strong pinning potential instead favors the $1$ sector and reverses the level ordering, bringing the states that were previously excited into the ground-state manifold.}
    \label{fig:binding_scheme}
\end{figure}

When two quasiholes are fused, realistic interactions lift the degeneracy between their fusion channels and energetically select a preferred one (see \Cref{fig:binding_scheme}). The Coulomb interaction naturally favors the $\psi$ channel~\cite{baraban2009numerical,xu2025dynamics,gattu2025molecular}. By contrast, a local electrostatic potential that repels electrons favors the $1$ channel, since the electron density near the fused pair is lower in the $1$ channel than in the $\psi$ channel~\cite{trung2025long,seesup}. Such an electrostatic potential models the effect of using a scanning tunneling microscope (STM) tip to trap quasiholes~\cite{papic2018imaging}. The competition between the electron-electron interaction and local electrostatic confinement therefore determines the fusion channel, with important experimental consequences.

To numerically demonstrate these competing effects, we use the toy Hamiltonian
\begin{align}
    \hat H_{\text{total}}&=h_0\hat V_{3}^{\mathrm{3bdy}}+\hat V_{\mathrm{Coulomb}}+h\hat V_{\mathrm{pins}}\label{eq:Hamiltonian},
\end{align}
where $\hat V_{3}^{\mathrm{3bdy}}$ is the three-body pseudopotential and is the model Hamiltonian for the Moore-Read state~\cite{simon2007generalized}, and $h$ controls the strength of the local pinning potential relative to the Coulomb interaction. $\hat V_{\text{pins}}$ in \Cref{eq:Hamiltonian} consists of $2n$ identical potential pins, each being a Gaussian potential profile with a small enough width ($\sigma=0.05$), centered at locations $\eta_1$, $\eta_2$,...,$\eta_{2n}$~\cite{seesup}. To induce a pair fusion, we set $\eta_1=\eta_2=0$, effectively fusing two quasiholes at the north pole of the sphere. Consequently, small $h$ favors fusion into $\psi$, while sufficiently large $h$ favors fusion into $1$. 
In the limit $h_0\to\infty$, the low-lying eigenstates of \Cref{eq:Hamiltonian} are expected to be well described by the MR fusion space of $2n$ quasiholes, each located at one pin. Thus, we use the HMC method to compute the matrix elements of only the last two terms of \Cref{eq:Hamiltonian} and diagonalize the resulting $2^{n-1}\times2^{n-1}$ matrix.




\begin{figure}[htp!]
\includegraphics[width=\columnwidth]{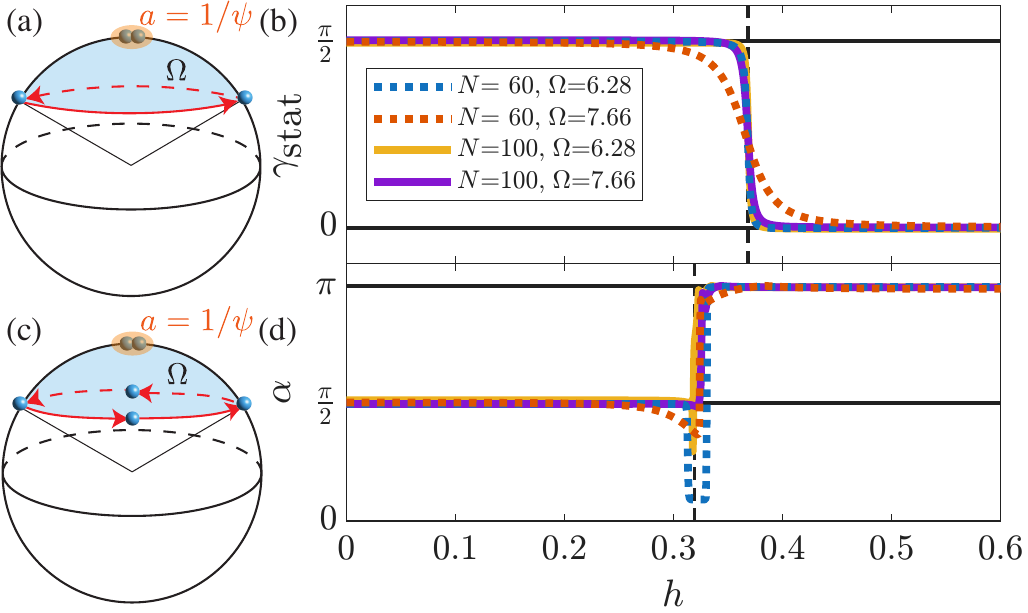}
\caption{\textbf{Numerical results for many-anyon braiding in the Moore-Read state.} Panels (a) and (b) show the four-quasihole geometry and the resulting statistical Berry phase $\gamma_{\mathrm{stat}}$, respectively. Panels (c) and (d) show the corresponding six-quasihole geometry and the relative eigenvalue phase $\alpha$ of the non-Abelian braiding matrix. The horizontal lines are the theoretical predictions~\cite{seesup}. In both geometries, two quasiholes are fused at the north pole, while the remaining quasiholes are braided in the presence of pinning potentials of strength $h$. Panels (b) and (d) share the same legend in panel (b).}
\label{fig:result_angles}
\end{figure}

We first examine the four-quasihole geometry in Fig.~\ref{fig:result_angles}(a), for which \new{two quasiholes are fused at the north pole, and the remaining two are swapped by a $\pi$-rotation at a fixed latitude}. The total Berry phase is obtained from the total angular momentum expectation value~\cite{umucalilar2018time,macaluso2019fusion,Prodan_Mapping_2009,trung2023spin}, that is $\gamma_{\mathrm{tot}}(\Omega)=\pi\left(\av{L_z}_{\Omega}-\av{L_z}_{\Omega=0}\right)$, with $\Omega$ being the solid angle enclosed by the braiding trajectories and $\av{L_z}=(S+1)\av{\cos\theta}$ in the lowest Landau level, where $\theta$ is the polar angle coordinate. \new{$\av{L_z}_{\Omega=0}$ denotes the expectation value of the angular momentum in the reference configuration with all quasiholes located at the north pole.} Subtracting the geometric contributions gives $\gamma_{\mathrm{stat}}=\gamma_{\mathrm{tot}}-\left(N/4+1/8\right)\Omega+\pi/2$~\cite{arovas1984fractional,seesup}. \new{As shown in \Cref{fig:result_angles}(b), $\gamma_{\mathrm{stat}}$ approaches $\pi/2$ as $h\rightarrow0$ and $0$ at large $h$. These limiting values precisely agree with the predictions established from the fusion-channel analysis: Coulomb-dominated interactions at small $h$ select the $\psi$ channel, yielding $\gamma_{\mathrm{stat}}=\pi/2$, whereas strong pinning selects the vacuum $1$ channel, yielding $\gamma_{\mathrm{stat}}=0$ (see \Cref{fig:binding_scheme})~\cite{trung2025long}. This agreement demonstrates that tuning the pinning strength switches the fusion channel of the fused pair.}

We next consider the six-quasihole geometry shown in \Cref{fig:result_angles}(c), \new{in which the four quasiholes away from the north pole are braided by a $\pi/2$ rotation.} After the north-pole pair is fused, the interaction-selected low-energy manifold is two-dimensional, so the non-Abelian braiding acts as a $2\times2$ matrix. Figure~\ref{fig:result_angles}(d) shows the relative eigenvalue phase
\begin{equation}
\label{eq:eigenvalue phase}
    \alpha=\text{Arg}\left(\frac{\lambda_1}{\lambda_2}\right),
\end{equation}
where $\lambda_1$ and $\lambda_2$ are the two eigenvalues of the braiding matrix. The phase $\alpha$ approaches $\pi/2$ as $h\to0$, when the north-pole pair occupies the $\psi$ fusion channel, and $\pi$ for large $h$, when the pair occupies the vacuum $1$ channel, \new{in agreement with the Ising-anyon predictions~\cite{seesup}.} 

\new{The abrupt changes of $\gamma_{\text{stat}}$ and $\alpha$ in \Cref{fig:result_angles} show that electrostatic potentials can affect braiding far beyond small geometric corrections. The pinning strength $h$ determines whether the fused pair is in the $1$ or the $\psi$ channel, and hence whether the remaining quasiholes are braided in the even or the odd topological sector; tuning $h$ thus flips this even-odd effect~\cite{seesup}, which experiments must account for. These results go beyond a numerical verification of the Ising CFT: the CFT fixes the braid once the fusion channel is specified, but cannot determine which channel is energetically selected under a generic Hamiltonian.} 

\newsect{Discussion}
In this work, we employ the HMC method~\cite{wang2026hybrid} to investigate the non-Abelian braiding of up to six quasiholes. This capability enables high-precision evaluation of many-anyon braiding matrices and reveals that braiding protocols based on global rotations are substantially more robust against finite-size effects than conventional two-anyon exchange schemes. \new{From the finite-separation dependence of the braiding matrix, we further extract a non-Abelian characteristic length $\xi\approx2.8\,\ell_B$, consistent with previous estimates for MR quasiholes.} We also show that electron-electron interactions and local trapping potentials can modify the braiding matrices by selecting the fusion channel of distant quasiholes, which directly reflects the long-range entanglement of non-Abelian states and makes control of the global fusion state essential in experiments. Although our calculations use model MR wavefunctions rather than exact eigenstates of realistic Hamiltonians, we expect the universal topological content of the braiding to be preserved as long as the realistic system remains adiabatically connected to the model Hamiltonian without closing the bulk gap; in this case, projecting realistic interactions onto the model quasihole manifold captures their leading-order effects~\cite{Wan2008,Tserkovnyak2003,Prodan_Mapping_2009}. Given numerical evidence that the $\nu=5/2$ ground state of a microscopic model is described by the MR state for sufficiently sharp confinement~\cite{Wan2008}, we expect the physics discussed here to be observable, at least qualitatively, in experiments.

The study of non-Abelian particles could be even more interesting in generic Chern bands realized in 2D quantum materials without external magnetic field~\cite{Cai2023_signature_fqah_mote2,Park2023_observation_fqah_mote2,Zeng2023_thermo_evidence_fqah_mote2,Xu2023_Observation_FQAH_tMote2,Lu2024_FQAH_multilayer_graphene,luExtended2025,park2025observation}, where fusion energetics is expected to be further enriched by fluctuating quantum geometric tensor. Most such systems can be well approximated by ideal flat bands~\cite{wangExact2021,pengEmergence2026,liu2015characterization,bernevig2025fractional,jaworowski2019characterization}, where many-body model wavefunctions can also be constructed analytically. The HMC methods can thus be easily generalized to a large family of Chern bands, especially in moir\'e systems, where the Moore-Read topological phase can potentially be realized~\cite{quantumNiu2025,nonabelianLiu2025}.

Along the direction of study into microscopic properties of non-Abelian anyons in FQH systems, it is important to note that the HMC method, as discussed here, can be 
generally applied to anyon models with more intricate fusion rules such as Fibonacci anyons, which in principle support universal topological quantum computation. In the FQH system, the model wavefunction for the Fibonacci anyons is known, but the literature on their braiding properties is more scarce~\cite{read1999beyond,wu2014braiding}. 
The HMC method is a promising tool for this problem because its computational advantages are not tied to a particular trial state, and it can be generalized whenever suitable many-quasihole wavefunctions and their gradients are available. With appropriate model wavefunctions, the method we show here could provide a practical route to computing the many-anyon braiding matrices and exploring the effects of interaction and confinement in candidate platforms for universal topological quantum computation.


\textit{Acknowledgments} --
We thank Steven H. Simon and Dung Xuan Nguyen for the constructive comments on the manuscript and for the suggested references. TTW, ML, and ZYM acknowledge the support from the Research Grants Council (RGC) of Hong Kong (Project Nos. AoE/P701/20, C7037-22GF,
17301924, 17301725, 17302026 and AoE/P-604/25-R), the ANR/RGC Joint Research Scheme sponsored by RGC of Hong Kong and the French National Research Agency (Project No. A\_HKU703/22), and the State Key Laboratory of Optical Quantum Materials at HKU. We thank the HPC2021 system under the Information Technology Services at the University of Hong Kong~\cite{hpc2021}, as well as the Beijing Paratera Tech Corp., Ltd~\cite{paratera} for providing HPC resources that have contributed to the research results reported within this paper. This work at Nanyang Technological University, Singapore, is supported by the Singapore Ministry of Education (MOE) Academic Research Fund Tier 1 Grant (No. RG156/24), Singapore Ministry of Education (MOE) Academic Research Fund Tier 3 Grant (No. MOE-MOET32023-0003) “Quantum Geometric Advantage”, and Singapore Ministry of Education (MOE) Academic Research Fund Tier 2 Grant (No. MOE-T2EP50124-0017).

\bibliographystyle{longapsrev4-2}
\bibliography{ref}

\end{document}


\title{Supplemental material for ``Many-Anyon Braiding in Non-Abelian Fractional Quantum Hall Effect with Hybrid Monte Carlo Simulation''}

\author{Ting-Tung Wang}
\thanks{These authors contribute equally.}
\affiliation{Department of Physics and HK Institute of Quantum Science \& Technology, The University of Hong Kong, Pokfulam Road,  Hong Kong SAR, China}
\affiliation{State Key Laboratory of Optical Quantum Materials, The University of Hong Kong, Pokfulam Road,  Hong Kong SAR, China}

\author{Ha Quang Trung}
\thanks{These authors contribute equally.}
\affiliation{Division of Physics and Applied Physics, Nanyang Technological University, Singapore 637371, Singapore}

\author{Qianhui Xu}
\affiliation{Division of Physics and Applied Physics, Nanyang Technological University, Singapore 637371, Singapore}

\author{Min Long}
\affiliation{Department of Physics and HK Institute of Quantum Science \& Technology, The University of Hong Kong, Pokfulam Road,  Hong Kong SAR, China}
\affiliation{State Key Laboratory of Optical Quantum Materials, The University of Hong Kong, Pokfulam Road,  Hong Kong SAR, China}

\author{Bo Yang}
\email{yang.bo@ntu.edu.sg}
\affiliation{Division of Physics and Applied Physics, Nanyang Technological University, Singapore 637371, Singapore}

\author{Zi Yang Meng}
\email{zymeng@hku.hk}
\affiliation{Department of Physics and HK Institute of Quantum Science \& Technology, The University of Hong Kong, Pokfulam Road,  Hong Kong SAR, China}
\affiliation{State Key Laboratory of Optical Quantum Materials, The University of Hong Kong, Pokfulam Road,  Hong Kong SAR, China}

\date{\today}

\makeatletter
\renewcommand{\theequation}{S\arabic{equation}}
\renewcommand{\thefigure}{S\arabic{figure}}
\setcounter{secnumdepth}{3}

\maketitle
\section{The Ising anyon model}\label{Sec:Ising_anyon}
The Ising anyon model contains three anyon species (topological charges) 1, $\psi$, and $\sigma$. The fusion rules, F-matrices, and R-matrices are given in \Cref{table:ising}, which completely define the model. We are interested in states with $2n$ $\sigma$-anyons, which could have a total topological charge of either 1 or $\psi$. (The total topological charge of a collection of anyons is the resulting anyon obtained from repeatedly fusing them pairwise until only one remains.) States with different topological charges reside in different topological sectors, which are referred to as the even sector (total charge 1) and the odd sector (total charge $\psi$). To explicitly write down a given state, one must specify the ``history" of how these $2n$ anyons are created pairwise from the total charge. This can be done by writing down the tree diagram, and a common choice of basis is as follows
\begin{equation}
    \label{eq:2n sigmas gen}
    \vcenter{\hbox{\includegraphics[width=0.25\linewidth]{trees/2nsigmasgen}}}
\end{equation}
where each of $a_0,a_1,...,a_{n-1}$ can be either 1 or $\psi$. The tree diagram can be read as a sequence of pair creation processes with time going upward. The total charge $a_0$ is fixed by the odd or even sector, and the remaining $a_i$'s have no constraint. Thus, within each sector, we see that there are $2^{n-1}$ possible states, giving rise to the non-Abelian degeneracy. Note that there are many ways that $2n$ $\sigma$-anyons can be created by consecutive pair creations starting from $a_0$, and \Cref{eq:2n sigmas gen} presents a specific choice of basis for the $2^{n-1}$-fold degenerate non-Abelian fusion space. Different choices of basis are related by the F-matrices, whose consistency is ensured by the so-called pentagon relation.

\begin{table}
    \begin{tabular}{|l|c|}
    \hline\
    Topological charges & 1,$\psi$, and $\sigma$\\
    \hline
    Fusion rules & \makecell{$\sigma\times\sigma=1+\psi$ \\ $\psi\times\sigma=\sigma$ \\ $\psi\times\psi=1$}\\
    \hline
    F-matrices & \makecell{$F_{\psi\sigma\psi}^\sigma=F_{\sigma\psi\sigma}^\psi=-1$\\%
                            $F_{\psi\psi\psi}^\psi=F_{\psi\sigma\sigma}^\psi=F_{\sigma\sigma\psi}^\psi=1$\\
                            $F_{\sigma\psi\psi}^\sigma=F_{\psi\psi\sigma}^\sigma=1$\\
                            $F_{\sigma\psi\sigma}^1=F_{\psi\sigma\sigma}^1=F_{\sigma\sigma\psi}^1=1$\\
                            $F_{\sigma\sigma\sigma}^\sigma=\frac{1}{\sqrt2}\begin{pmatrix}1&1\\1&-1\end{pmatrix}$}\\
    \hline 
    R-matrices & \makecell{$R^1_{\psi\psi}=-1$\\$R^\sigma_{\psi\sigma}=R^\sigma_{\sigma\psi}=-i$\\$R^1_{\sigma\sigma}=e^{-i\pi/8}$\\$R^\psi_{\sigma\sigma}=e^{i3\pi/8}$}\\
    \hline 
    \end{tabular}
    \caption{Fusion rules, F-matrices, and R-matrices of the Ising anyon model.}
    \label{table:ising}
\end{table}

The exchange of two anyons is described at the most primitive level by the R-matrix, defined as
\begin{equation}
    \label{eq:R-matrix}
    \vcenter{\hbox{\includegraphics[width=0.05\linewidth]{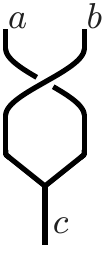}}}=R^c_{ab}\vcenter{\hbox{\includegraphics[width=0.05\linewidth]{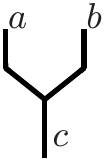}}}
\end{equation}
which describes the outcome of exchanging the positions of any two anyons formed from the same parent. A direct application of the R-matrix is in describing the exchange of two $\sigma$-anyons, which have different outcomes in the odd and even sectors:
\begin{align}
\vcenter{\hbox{\includegraphics[width=0.05\linewidth]{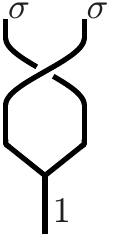}}}=R_{\sigma\sigma}^1\vcenter{\hbox{\includegraphics[width=0.05\linewidth]{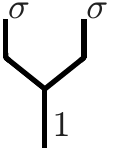}}}=e^{-i\pi/8}\vcenter{\hbox{\includegraphics[width=0.05\linewidth]{trees/sigmasigma1.pdf}}},\label{eq:sigmasigma1}\\
    \vcenter{\hbox{\includegraphics[width=0.05\linewidth]{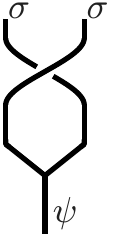}}}=R_{\sigma\sigma}^\psi\vcenter{\hbox{\includegraphics[width=0.05\linewidth]{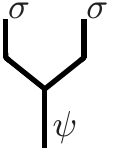}}}=e^{i3\pi/8}\vcenter{\hbox{\includegraphics[width=0.05\linewidth]{trees/sigmasigmapsi.pdf}}}\label{eq:sigmasigmapsi}.
\end{align}
We thus see that a system with only two $\sigma$ anyons is Abelian, with fusion space dimension $2^{1-1}=1$. However, the non-Abelian property of the anyon model manifests in the different scalar braiding phase in the different sectors. This is often termed the ``even-odd" effect.

The even-odd effect applies not only to the Abelian braiding of two anyons, but also applies generally to the braiding matrices for four or more $\sigma$'s. As an example, let us consider the case of four $\sigma$'s, for which the basis in the even and odd sectors can be written as
\begin{equation}
\label{eq:4sigmas}
    \vcenter{\hbox{\includegraphics[width=0.125\linewidth]{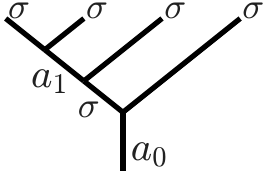}}},\hspace{11pt}a_0=\begin{cases}
        1, &\text{ even}\\
        \psi, &\text{ odd}
    \end{cases}.
\end{equation}
We define the braiding ``generators", each as the result of exchanging a pair of $\sigma$'s in the consecutive ``slots" on the tree diagrams. In the case of four $\sigma$'s, there are three such generators:
\begin{align}
    \vcenter{\hbox{\includegraphics[width=0.125\linewidth]{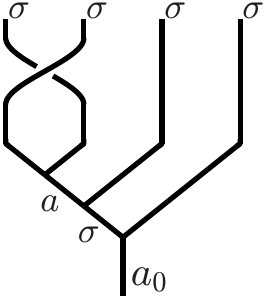}}}=\sum_bB_{ab}^{(12)}\vcenter{\hbox{\includegraphics[width=0.125\linewidth]{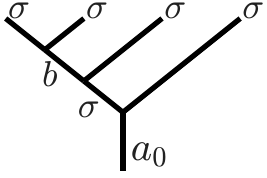}}},\label{B12def}\\
    \vcenter{\hbox{\includegraphics[width=0.125\linewidth]{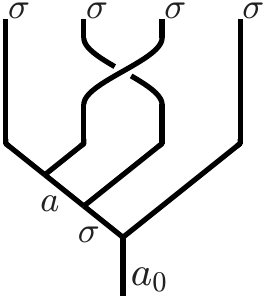}}}=\sum_bB_{ab}^{(23)}\vcenter{\hbox{\includegraphics[width=0.125\linewidth]{trees/4sigmasb.pdf}}},\label{B23def}\\
    \vcenter{\hbox{\includegraphics[width=0.125\linewidth]{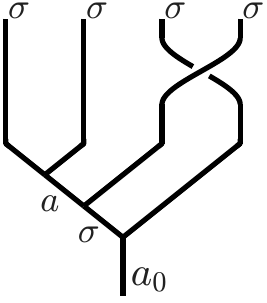}}}=\sum_bB_{ab}^{(34)}\vcenter{\hbox{\includegraphics[width=0.125\linewidth]{trees/4sigmasb.pdf}}}.\label{B34def}
\end{align}
By picking an ordering for the chosen basis, every state in a given fusion space (with a fixed even or odd sector) can be written as a spinor vector with each component representing the corresponding coefficient. Then, each term $B^{(12)}$, $B^{(23)}$, and $B^{(34)}$ can be written as a $2\times2$ matrix. If we choose the spinor such that $(c_1,c_2)^T$ represents the state
\begin{equation}
    \label{eq:4 sigma spinor}
    c_1\vcenter{\hbox{\includegraphics[width=0.125\linewidth]{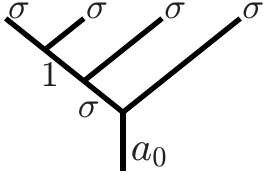}}}+c_2\vcenter{\hbox{\includegraphics[width=0.125\linewidth]{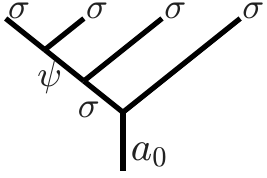}}}.
\end{equation}
Then the three generators in \Cref{B12def,B23def,B34def} can be written as
\begin{align}
    B^{(12)}&=e^{-i\pi/8}\begin{pmatrix}1&0\\0&i\end{pmatrix}\label{B12matrix},\\
    B^{(23)}&=\frac{e^{-i\pi/8}}{2}\begin{pmatrix}1+i&1-i\\1-i&1+i\end{pmatrix}\label{B23matrix},\\
    B^{(34)}&=\begin{cases}
        e^{-i\pi/8}\begin{pmatrix}1&0\\0&i\end{pmatrix},&\text{even sector}\\
        e^{-i\pi/8}\begin{pmatrix}i&0\\0&1\end{pmatrix},&\text{odd sector}
    \end{cases}\label{B34matrix}.
\end{align}
Among the three generators, only $B^{(12)}$ and $B^{(23)}$ are identical in both the even and odd sectors. The braiding matrix of any general braiding scheme can be written as a product of these three matrices (hence the term \emph{generator}). Since $B^{(34)}$ takes different values in the two sectors, we see that if a braiding scheme involves all four quasiholes, then in general the outcome will be dependent on the sector in which the four quasiholes reside.

This discussion generalizes to the fusion space of $2n$ $\sigma$-anyons, where the braiding generator $B^{(i,i+1)}$ is defined as the result of exchanging the anyons in the $i^{\text{th}}$ and $(i+1)^{\text{th}}$ slot, for $i=1,2,...,2n-1$. These generators provide a way to quickly compute the result of any braiding scheme, up to a basis transformation.

\section{First-quantized wavefunction}\label{Sec:SM_wavefunction}
The Moore-Read state with $2n$ quasiholes on a sphere with the usual stereographic projection is written as
\begin{widetext}
\begin{equation}
\label{eq:MR quasihole}
    \Psi_{\mathrm{MR}}(\{z_i\},\{\eta_k\})
    =
    \mathrm{Pf}\!\left(\frac{\prod_{k=1}^n(z_i-\eta_k)(z_j-\eta_{k+n})+(i\leftrightarrow j)}{z_i-z_j}\right)
    \prod_{i<j}(z_i-z_j)^2
    \prod_i\left (\frac{1}{1+|z_i|^2}\right )^S.
\end{equation}    
\end{widetext}
Here $z_i=\tan(\theta_i/2)e^{i\varphi_i}$ is the holomorphic variable parameterizing the position of the $i^{\text{th}}$ electron on the sphere (with $\theta_i$ and $\varphi_i$ being its polar and azimuthal coordinates, respectively), and $\eta_k$ similarly parametrizes the position of the $k^{\text{th}}$ quasihole. $2S=2N+n-3$ is the total number of flux quanta, which is an integer due to Dirac quantization~\cite{haldane1983fractional,greiter2011landau}. Throughout this paper, we work with the unit $\hbar=c=eB=1$ (here $B$ is the value of the uniform magnetic field normal to the surface of the 2DEG). The Pfaffian factor encodes a paired structure of composite fermions and is responsible for the emergence of Ising-type non-Abelian statistics. Quasiholes in this state are associated with insertions of Ising spin fields $\sigma(\eta_a)$ at quasihole positions $\eta_a$, together with the appropriate charged vertex operators. The resulting many-body wavefunctions are conformal blocks of the Ising conformal field theory combined with a charged boson sector.

Permuting the variables $(\eta_1,\eta_2,...,\eta_{2n})$ results in $(2n)!$ different wavefunctions of the form in \Cref{eq:MR quasihole}. Of these, one can only find $2^{n-1}$ linearly independent states \cite{nayak19962n}. This is because the MR quasiholes realize the Ising anyon model. The electric charge of a quasihole is defined as the total deviation from the average electron density. This electric charge is often described by a $U(1)$ part coupled to the Ising theory, which we will not discuss in detail in this paper. We note that the $\sigma$-anyon carries charge $e/4$ while the $1$-anyon and $\psi$-anyon each carry charge $e/2$. The $1$-anyon and $\psi$-anyon in the MR state carry the same total electric charge but have different internal structures (charge density distributions).

The fundamental fusion rule of the quasiholes $\sigma(\eta_\alpha)$is
\begin{equation}
    \sigma \times \sigma = 1 + \psi ,
\end{equation}
where $1$ denotes the vacuum sector and $\psi$ is the neutral fermion sector. Thus, bringing two quasiholes together does not lead to a unique outcome: the pair may fuse either to $1$ or to $\psi$. This fusion ambiguity produces a multi-dimensional space of nearly degenerate quasihole states. Braiding operations act within this space through non-commuting unitary matrices, rather than by simple Abelian phases. Consequently, the braiding properties depend not only on the geometric exchange path, but also on the fusion channel occupied by the quasiholes.

In the four-quasihole case, the Moore--Read state has a two-dimensional space of conformal blocks. A convenient basis is obtained by specifying the fusion channel of a chosen pair of quasiholes, for example
\begin{equation}
    \left[ (\sigma_1 \sigma_2)_a (\sigma_3 \sigma_4)_a \right]_{1},
    \qquad
    a \in \{1,\psi\},
\end{equation}
assuming an overall trivial topological charge. On this basis, the two possible fusion outcomes of the first pair label the two non-Abelian states. Braiding neighboring quasiholes changes the relative phases and, for exchanges involving different fusion pairings, mixes the two basis states through the Ising $F$ and $R$ matrices given in \Cref{table:ising}. This makes the four-quasihole sector the minimal setting in which the fusion dependence of Moore--Read braiding can be observed explicitly~\cite{Tserkovnyak2003,wang2026hybrid}.

For six quasiholes $(n=3)$, the fusion space is enlarged, supporting four independent conformal blocks before imposing any additional global constraint. A basis may be chosen by sequentially fusing quasiholes pairwise or along a fusion tree, with intermediate channels taking values in $1$ or $\psi$ subject to consistency with the total topological charge. Compared with the four-quasihole case, the six-quasihole sector allows a richer set of fusion pathways and therefore a larger representation of the braid group. Exchanges of quasiholes act by local $R$ moves in a given fusion channel, combined with $F$ moves whenever the exchange is not diagonal in the chosen fusion basis.

The effect of quasihole fusion on braiding can therefore be understood as a consequence of the nontrivial topology of the Moore--Read quasihole Hilbert space. When two quasiholes are adiabatically fused, the resulting state is projected into either the $1$ or $\psi$ channel, and subsequent braiding operations act differently depending on this projection. In particular, braids that are diagonal in one fusion basis may become non-diagonal after a change of fusion basis, leading to channel-dependent interference and state mixing. For systems with four to six quasiholes, this interplay between fusion and braiding provides a direct probe of the Ising anyon structure underlying the Moore--Read phase~\cite{bonderson2006detecting,Bonderson2011}.

\label{sec:theoretical}
\begin{figure}
    \centering
    \includegraphics[width=0.5\linewidth]{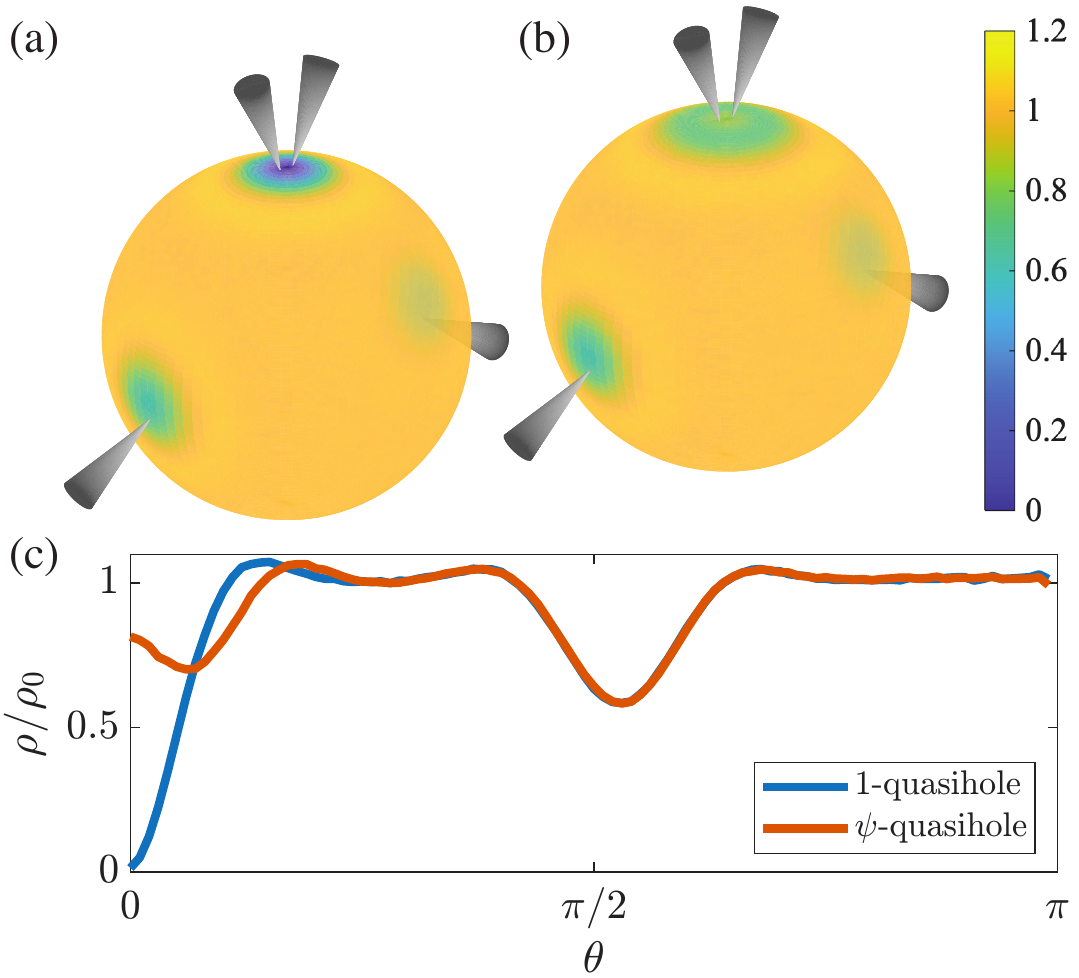}
    \caption{The electron density on the sphere for the Moore-Read state with 60 electrons and 4 quasiholes, 2 at the north pole and the other 2 located at the same latitude. The quasiholes at the north pole fuse and form a (a) $1$-anyon or (b) $\psi$-anyon. (c) Comparison of the electron density along the longitudinal direction, cutting through one of the lower quasiholes, for the two states.}
    \label{fig:MR quasihole density}
\end{figure}
\section{Classical and rotational braiding scheme}
The theoretical braiding matrices for the two schemes can be obtained directly by resolving the corresponding Ising-anyon fusion trees using the $F$- and $R$-moves. We use the basis in which the intermediate fusion channel of a pair of $\sigma$ anyons is $a=1,\psi$. For the classical braiding scheme, the exchanged quasiholes originate from the same parent branch of the fusion tree, so that the crossing is resolved directly by the $R$-matrix. The resulting braiding matrix is
\begin{equation}
\label{eq:classical}
\vcenter{\hbox{\includegraphics[width=0.125\linewidth]{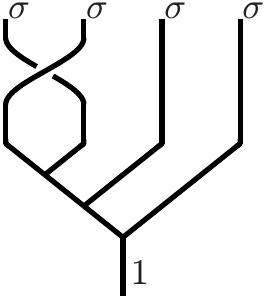}}}\;\;\;
B_{\mathrm{cl}}
=B^{(12)}
=e^{-i\pi/8}
\begin{pmatrix}
1 & 0\\
0 & i
\end{pmatrix}.
\end{equation}
Its two eigenvalues therefore differ by a phase $e^{i\alpha}=\lambda_1/\lambda_2=-i$, giving the theoretical prediction $\alpha=-\pi/2$. Note that since the four $\sigma$'s are identical, the braiding process can also be described by $B^{(23)}$. The exact outcome depends on the choice of basis, i.e., the identification of the four identical $\sigma$'s in the system with the four branches in the fusion tree. Indeed, $B^{(12)}$ and $B^{(23)}$ are related by a basis transformation represented by $F_{\sigma\sigma\sigma}^\sigma$ (see \Cref{table:ising}). Thus, in this paper we focus on the basis-independent comparison between two braiding matrices, that is, the relative phase $\alpha$ of the two eigenvalues. This relative phase is also insensitive to the overall scalar $U(1)$ part of the total Berry matrix.

\begin{figure}
    \centering
    \includegraphics[width=0.8\linewidth]{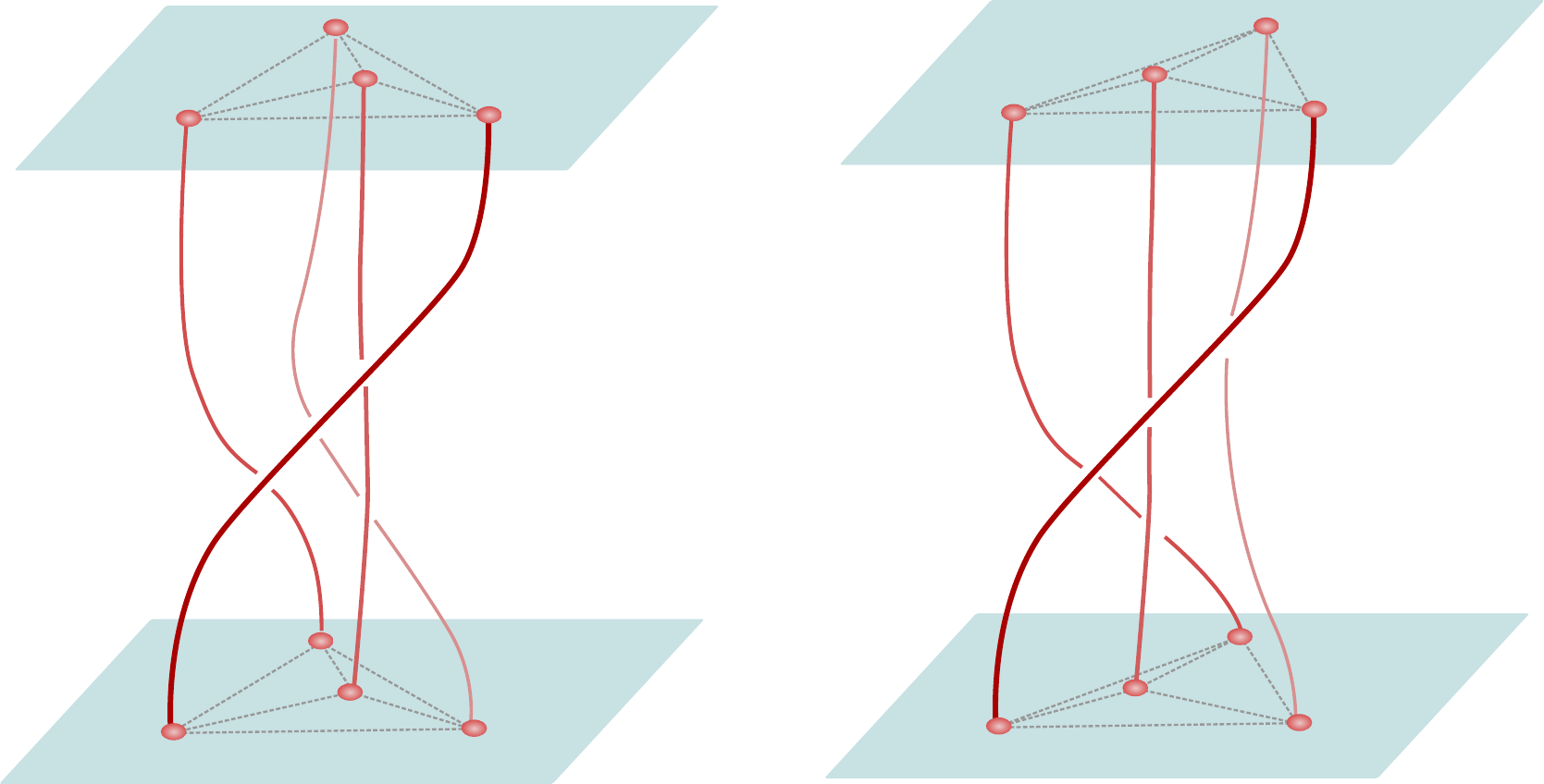}
    \caption{Two different representations of the same rotational braiding scheme involving four anyons and a $2\pi/3$-rotation. The left panel gives the braiding matrix in \Cref{eq:rotational 1} while the right panel gives \Cref{eq:rotational 2}.}
    \label{fig:rotation representations}
\end{figure}

\begin{figure}
    \centering
    \includegraphics[width=\linewidth]{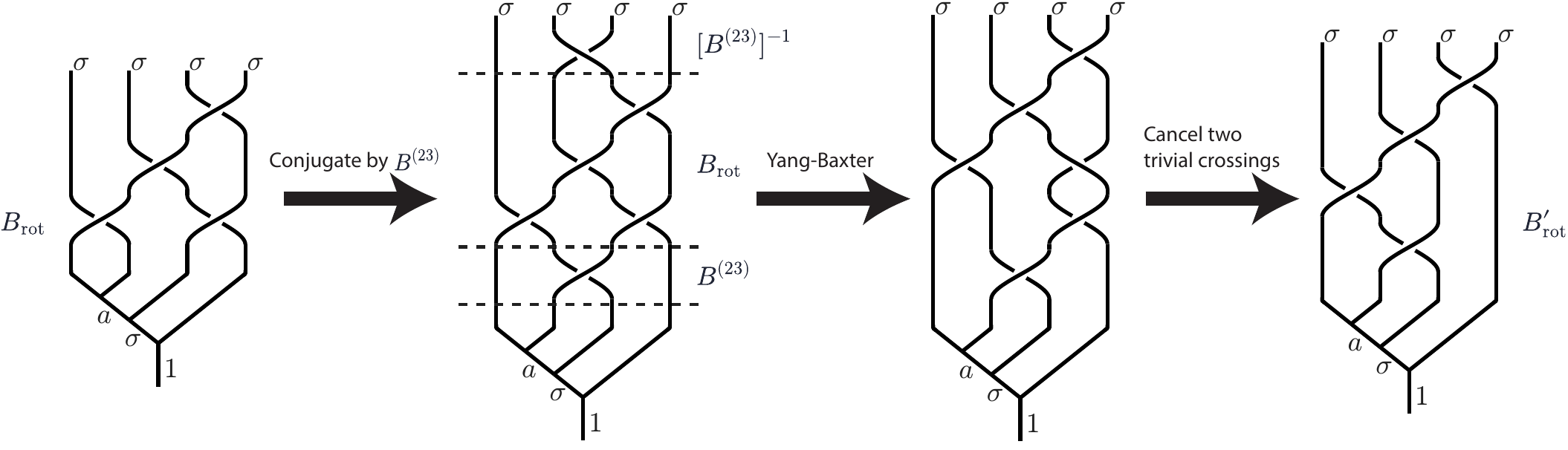}
    \caption{Equivalence between the two tree diagrams describing the two representations in \Cref{fig:rotation representations}.}
    \label{fig:equivalence}
\end{figure}

For the rotational braiding scheme, the $2\pi/3$ collective rotation corresponds to a sequence of elementary exchanges in the fusion tree. Resolving these crossings from bottom to top gives
\begin{equation}
\label{eq:rotational 1}
\vcenter{\hbox{\includegraphics[width=0.125\linewidth]{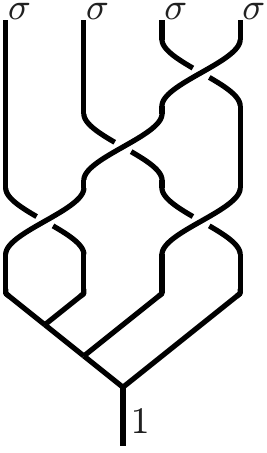}}}\;\;\;
B_{\mathrm{rot}}
=B^{(34)}B^{(23)}B^{(12)}
B^{(34)}
=\frac{1}{2}
\begin{pmatrix}
1-i & 1+i\\
1-i & -1-i
\end{pmatrix}.
\end{equation}
The two eigenvalues can be written as $e^{-i\pi/6}$ and $e^{-i5\pi/6}$, whose ratio gives $e^{i\alpha}=e^{-i2\pi/3}$ and hence $\alpha=-2\pi/3$. Although the explicit matrix representation depends on the fusion basis and an overall Abelian phase, the relative eigenvalue phase $\alpha$ provides the basis-independent quantity compared with our numerical results. Note that similar to the discussion in the classical scheme, the braiding matrix describing the rotational scheme is also basis-dependent. Since the braiding involves many anyons, different representations of the same braiding scheme (see \Cref{fig:rotation representations} for two examples) may result in a completely different sequence of pair exchanges and thus a different braiding matrix. For example, the left panel of \Cref{fig:rotation representations} gives \Cref{eq:rotational 1} while the right panel gives
\begin{equation}
    \label{eq:rotational 2}
    \vcenter{\hbox{\includegraphics[width=0.125\linewidth]{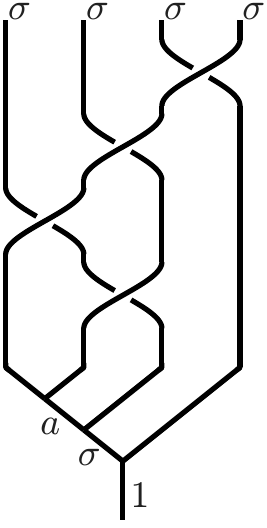}}}\;\;\;
    B_{\text{rot}}'= B^{(34)}B^{(23)}B^{(12)}B^{(23)}=\frac{1}{2}
    \begin{pmatrix}
    1-i & 1-i\\
    1+i & -1-i
    \end{pmatrix}.
\end{equation}
However, since this difference merely arises from a basis (unitary) transformation, the two matrices must necessarily give the same eigenvalue phase ratio $\alpha$. (Indeed, that can be easily seen since $B_{\text{rot}}'$ is the transpose of $B_{\text{rot}}$.) The basis transformation can be seen explicitly as in \Cref{fig:equivalence}, which gives:
\begin{equation}
    \label{eq:rotation braiding matrix equivalence}
    B_{\text{rot}}'=\left[B^{(23)}\right]^{-1}B_{\text{rot}}B^{(23)}.
\end{equation}

\section{Finite-size effect and origin of the robustness of the rotational scheme}
A major factor contributing to the finite-size effect in Berry matrix calculation is that the model wavefunctions for the different fusion channels within the fusion space are only orthogonal in the thermodynamic limit~\cite{Bonderson2011}. This can also explain the robustness of the rotational braiding scheme as follows.

Suppose the $d$-dimensional fusion space $\mathcal H$ is spanned by $|\psi_1\rangle,|\psi_2\rangle,...,|\psi_d\rangle$. We vary this fusion space in some parameter space parametrized by a fictitious time $t$ to get $\mathcal H(t)=\text{span}\left\{|\psi_1(t)\rangle,...,|\psi_d(t)\rangle\right\}$ and we assert that $\mathcal H(T)=\mathcal H(0)$ for some time $T$. The $(i,j)$-entry of the Berry connection tensor describing this process is defined as
\begin{equation}
    \label{eq:Berry connection}
    A_{i,j}=i\langle\psi_i|\frac{d}{dt}|\psi_j\rangle
\end{equation}
In a finite system, each basis state contains a small correction coming from the orthonormalization process, resulting in the new basis $|\tilde\psi_i\rangle=|\psi_i\rangle+\varepsilon_i|\delta \psi_i\rangle$. 
The coefficients $\varepsilon_i\to0$ in the thermodynamic limit. \Cref{eq:Berry connection} thus contains correction terms of the form:
\begin{equation}
    \label{eq:correction}
    \tilde A_{i,j}=i\langle\psi_i|\frac{d}{dt}|\psi_j\rangle + i\varepsilon_j\langle\delta\psi_i|\frac{d}{dt}|\psi_j\rangle +i\varepsilon_j\langle\psi_i|\frac{d}{dt}|\delta\psi_j\rangle + i\langle\psi_i|\delta\psi_j\rangle\frac{d\varepsilon_j}{dt}+\mathcal O(\varepsilon^2),
\end{equation}
which in general add corrections to the Berry matrix in the finite system. However, if the closed loop is generated by a global rotation such that:
\begin{align}
|\psi_i(t)\rangle&=e^{i\omega t\hat L_z}|\psi_i(0)\rangle\\
|\delta\psi_i(t)\rangle&=e^{i\omega t\hat L_z}|\delta\psi_i(0)\rangle
\end{align}
then by symmetry $\varepsilon_j$ is constant and the term with $d\varepsilon_j/dt$ in \Cref{eq:correction} vanishes. In contrast, in the classical scheme this term contributes a path-dependent correction to the Berry matrix.

Note that the advantage of the rotational braiding scheme compared to the classical scheme is a result of the CFT wavefunctions being non-orthogonal in the finite systems. This is merely an effect of choosing a basis and thus does not reflect the physics of anyons, in particular the effects of their finite size. Thus, we see that in both braiding schemes the monodromy approaches theoretical values as the anyon separation is increased with the same characteristic length scale.

\section{Spontaneous fusion channel selection}
We consider a general fusion space consisting of $2n$ $\sigma$-anyons as shown in \Cref{eq:2n sigmas fused two}. For simplicity, we will fix the even sector by fixing $a_0=1$, but the same arguments will apply to the odd sector ($a_0=\psi$), only with some qualitatively different results. Suppose we consider a braiding process that involves only $2n-2$ anyons. If all anyons are well-separated in real space, it is expected that the fusion space is exactly $2^{n-1}$-fold degenerate, and in general, the physics of the subspace of $2n-2$ anyons can be determined by taking a partial trace over the last two anyons. The computation of the partial trace, especially in the context of anyon braiding, has been extensively studied in the past~\cite{bonderson2006detecting,bonderson2009splitting,bonderson2008interferometry}, and a particularly detailed discussion can be found in Ref.~\cite{bonderson2008interferometry}.

Here, we consider the case where the remaining two ``background'' anyons are fused, in which case the tree diagram of the basis can be written as
\begin{equation}
    \label{eq:2n sigmas fused two}
    \vcenter{\hbox{\includegraphics[width=0.216\linewidth]{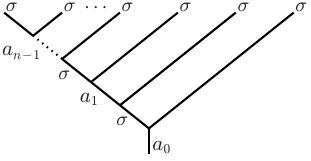}}}\rightarrow\vcenter{\hbox{\includegraphics[width=0.2125\linewidth]{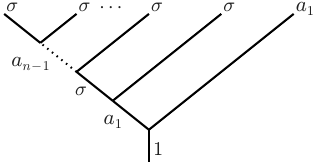}}}.
\end{equation}
In the MR state, the two states resulting from two different fusion channels $a_1=1$ or $\psi$ are only energetically degenerate when the underlying electron-electron interaction is the model three-body pseudopotential $\hat V_3^{3bdy}$~\cite{simon2007generalized}. When the electron-electron interaction is two-body, such is the case as the usual Coulomb interaction, there is a small but finite energy split in these fusion channels~\cite{baraban2009numerical,xu2025dynamics,gattu2025molecular}. Thus, the original $2^{n-1}$-fold degenerate fusion space is split into two halves, each of dimension $2^{n-2}$. The half with the lower energy, which we will refer to as the ground state manifold, is effectively a fusion space of $2n-2$ anyons, but with the total charge $a_1$, which is determined by whether the $1$ or $\psi$ fusion channel is energetically favoured by the underlying electron-electron interaction.

Besides the electron-electron interaction, the fusion channel selection of $a_1$ can also be influenced by a ``one-body'' electrostatic potential. This is because an electrostatic potential couples to the electron density, and we have seen from \Cref{fig:MR quasihole density} that the $1$-quasihole and $\psi$-quasihole of the MR state have different local density profiles. The individual effects of the two-body interaction and electrostatic potential can be summarized as follows:
\begin{itemize}
    \item The Coulomb interaction, $\hat V_{\text{Coulomb}}$, energetically favors the $\psi$-quasihole over the $1$-quasihole~\cite{baraban2009numerical,xu2025dynamics,gattu2025molecular}.
    \item A local electrostatic potential (which repels electrons) energetically favors the $1$-quasihole over the $\psi$-quasihole~\cite{trung2025long}. This is because in the small vicinity around the quasihole center, the $1$-quasihole has a smaller electron density compared to the $\psi$-quasihole, as seen in \Cref{fig:MR quasihole density} (c).
\end{itemize}
By making use of the contrasting effect of these two potentials, the fusion channel can thus be manually selected by tuning the parameters in an overall Hamiltonian of the form
\begin{equation}
    \label{eq:Hamiltonian}
    \hat H_{\text{total}}=h_0\hat V_{3}^{3bdy}+\hat V_{\mathrm{Coulomb}}+h\hat V_{\mathrm{pins}},
\end{equation}
where $\hat V_{\text{pins}}$ denotes an electrostatic potential consisting of several local potentials ``pins'', and $h$ controls the relative strength between the pinning and electron interaction. In this work, we choose the pinning potential such that each pin has a Gaussian profile of the form:
\begin{equation}
\label{eq:Gaussian potential}
    \hat V_{\mathrm{pins}}
    =
    \sum_{i=1}^{N}\sum_{k=1}^{2n}
    V_{\mathrm{pin}}(\mathbf r_i;\mathbf \eta_k),
    \qquad
    V_{\mathrm{pin}}(\mathbf r_i;\mathbf \eta_k)
    =
    \frac{1}{2\pi\sigma^2}
    \exp\!\left[
        -\frac{|\mathbf r_i-\mathbf \eta_k|^2}{2\sigma^2}
    \right],
\end{equation}
where each $\eta_k$ is the location of the center of the pin, and $\sigma$ is the width of the pin, which we choose to be 0.05 in our work. The fusion channel can be chosen by tuning the $h$ parameter in \Cref{eq:Hamiltonian}, with $h\to0$ favoring the $\psi$-fusion channel and $h\to\infty$ favouring the $1$-fusion channel. It will be shown by explicit computation that the braiding statistics of the subspace consisting of $2^{n-2}$ states are affected by the fusion channel selection. Two special cases with four and six anyons are described below.

\subsection{Four-anyon braiding}
\label{sec:theory_4_anyon}
We first look at a special case of the general discussion above, with $2n=4$ $\sigma$-quasiholes. After fusing two quasiholes, selecting a single fusion channel by \Cref{eq:Hamiltonian}, the remaining two quasiholes exhibit Abelian braiding statistics that follow either the even or odd sector, 
depending on the fusion outcome. 

To realize this effect, we employ a four-potential-pin setup where two potential pins are placed at the north pole, and the other two at the opposite ends on a circle of latitude. In a system with four MR quasiholes (MR state with two additional flux quanta), a charge-$e/2$ quasihole is pinned at the north pole, while a charge-$e/4$ quasihole is pinned at each of the other two potential pins. A braiding scheme is generated by rotating the sphere by $\pi$ around the $z$-axis, which is described by the following tree diagram
\begin{equation}
    \label{eq:4 anyon scheme}
    \vcenter{\hbox{\includegraphics[width=0.125\linewidth]{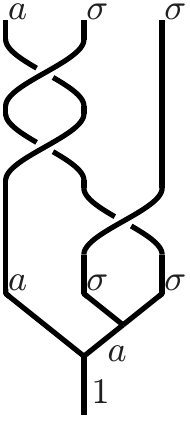}}}.
\end{equation}
The resulting braiding phase $\gamma_{\text{br}}$ depends on whether the fusion channel is even $(a=1)$ or odd $(a=\psi)$:
\begin{equation}
\label{eq:4 anyon theory}
    \gamma_{\text{br}}=\begin{cases}
        0,&\text{even}\\
        -\pi/2,&\text{odd}
    \end{cases}.
\end{equation}
Note that in the odd sector, the negative sign comes from the double braid of a $\sigma$ and a $\psi$. To extract only the mutual statistics $\gamma_{\text{stat}}$, which is the quantity shown in Fig. 1 in the main text, we take the negative of the value of the odd sector (the even-sector remains the same since any braiding with $1$-anyons is trivial). Thus, we obtain:
\begin{equation}
    \gamma_{\text{stat}}=\begin{cases}
        0,&\text{even}\\
        \pi/2,&\text{odd}
    \end{cases}.
\end{equation}
\subsection{Six-anyon braiding}
\label{sec:theory_6_anyon}
Among $2n=6$ quasiholes, if two quasiholes are fused into a given fusion channel, the non-Abelian braiding of the remaining four quasiholes will effectively obey the statistics of four quasiholes in either the even or odd sector, depending on the fusion channel. Similar to the four-quasihole case described above, we consider a braiding scheme with six potential pins: two potential pins at the north pole and the other four at the vertices of a square inscribing a circle of latitude. A rotation of the sphere by $\pi/2$ about the $z$-axis gives the following braiding scheme:
\begin{equation}
    \label{eq:6 anyon scheme}
    \vcenter{\hbox{\includegraphics[width=0.2\linewidth]{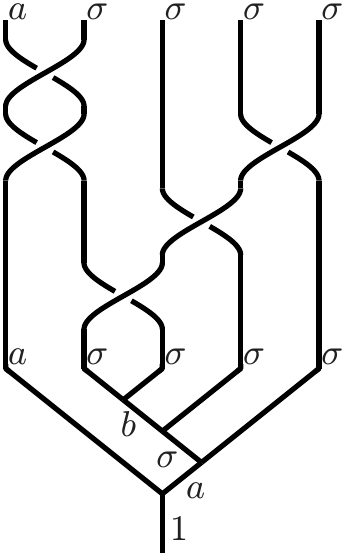}}}.
\end{equation}
Similar to the discussion above, the resulting braiding matrix $B$ depends on whether the selected fusion channel is even ($a=1$) or odd ($a=\psi$). In the even fusion sector, the braiding matrix is
\begin{equation}
    \label{eq:braiding matrix even}
    B^{\text{(even)}}=\frac{e^{-i\pi/8}}{\sqrt2}\begin{pmatrix}1&1\\1&-1\end{pmatrix} \text{, and }\alpha=\pi.
\end{equation}
And in the odd sector, the braiding matrix is
\begin{equation}
    \label{eq:braiding matrix odd}
    B^{\text{(odd)}}=-\frac{e^{3i\pi/8}}{\sqrt2}\begin{pmatrix}1&-1\\1&1\end{pmatrix} \text{, and }\alpha=\pi/2.
\end{equation}

\subsection{On the $U(1)$ phase}
In the four-anyon braiding case above, in order to extract the Abelian phase of the ground state, we subtract away $(N_\phi/8+1/4)\Omega - \pi/2$ from the total Berry phase, where $N_\phi=2S$ is the total number of fluxes through the sphere (note that $N_\phi=2N-1$ for a system with four MR quasiholes). The first term comes from the geometric phase, which consists of the Aharonov-Bohm phase and parallel transport of the quasihole spin along the surface of the sphere (both proportional to the enclosed solid angle)~\cite{trung2023spin}. The $\pi/2$ term, on the other hand, is not a geometric effect but arises from the electrical charge deficiency, i.e., the $U(1)$ part of the braiding phase. The origin of this term can be explained as follows.

Recall that moving a flux in a closed loop results in a total Berry phase that is proportional to the electron density within the enclosed loop~\cite{arovas1984fractional}:
\begin{equation}
    \label{arovas}
    \gamma = 2\pi \langle n\rangle_{\text{enc}} = -\frac{2\pi}{e}\langle\rho\rangle_{\text{enc}}
\end{equation}
where here $e$ denotes the magnitude of the electronic charge, $\langle n\rangle_{\text{enc}}$ denotes the average electron density within the enclosed loop, and $\langle\rho\rangle_{\text{enc}}$ denotes the average electrical charge within the enclosed loop.

In \cref{eq:4 anyon scheme} we see that a $\sigma$-quasihole is moved around an $a$-quasihole where $a$ can be either $1$ or $\psi$. Since a $\sigma$ is a half-flux, a factor of half should be added to the RHS of \Cref{arovas}. Since either $1$- or $\psi$-quasihole has electrical charge $e/2$, we see that the presence of $a$ inside a closed loop of $\sigma$ gives a phase shift:
\begin{equation}
    \label{eq:pi/2 phase shift}
    \Delta\gamma = \frac{1}{2}\times\left(-\frac{2\pi}{e}\delta\langle\rho\rangle_{\text{enc}}\right)=\frac{\pi}{e}\times\left(-\frac{e}{2}\right)=-\frac{\pi}{2}
\end{equation}
Thus, to extract only the statistical phase arising from exchanging two $\sigma$'s, we subtract -$\pi/2$ from the total Berry phase as well. Note that the results from computing the Berry phase by computing $\langle L_z\rangle$ differ from Ref.~\cite{arovas1984fractional} by an overall factor of $-1$ (see e.g. \cite{trung2023spin,umucalilar2018time}).

This $U(1)$ factor is also present in the six-anyon braiding in \Cref{eq:6 anyon scheme}, but since its contribution is an overall scalar factor, it does not affect the ratio of the eigenvalues of the braiding matrix, which is the quantity examined in this paper. Thus, we do not consider the $U(1)$ phase in the six-anyon case.

\section{Hybrid Monte-Carlo Method}\label{Sec:HMC}

This section offers a concise technical summary of our previous work~\cite{wang2026hybrid}. We refer the reader to the full details therein.

In the Hybrid Monte Carlo approach, the fractional quantum Hall sampling weight is written as a classical Boltzmann factor, $P(\{z_i\})=|\Psi(\{z_i\})|^{2}=\exp\left[{-V(\{z_i\})}\right]$. Consider the Moore-Read state on the disk as an example; one has
\begin{equation}
    V(\{z_i\})=-\log\left|\det\left(\frac{1}{z_i-z_j}\right)\right|-4\sum_{i<j}\log |z_i-z_j|+\frac{1}{2}\sum_i |z_i|^2 .
\end{equation}
With the presence of quasiholes, one simply changes the matrix inside the determinant according to~\Cref{eq:MR quasihole}. Then, one augments the electron coordinates with fictitious momenta $\{p_j\}$ and samples the extended Hamiltonian
\begin{equation}
\mathcal H(\{z_i\},\{p_i\})=V(\{z_i\})+\frac{1}{2}\sum_i |p_i|^2 ,
\enspace
\dot z_i=p_i,
\enspace
\dot p_i=-\frac{\partial V}{\partial z_i}.
\end{equation}

The Hamiltonian dynamics generates global updates for all particle coordinates while preserving the energy $\mathcal H$; in practice, it is integrated numerically by a symplectic leapfrog scheme and corrected by the Metropolis probability
\begin{equation}
r=\min\left \{1,\exp\left(\mathcal H_{\rm old}-\mathcal H_{\rm new}\right)\right \},
\end{equation}
so that detailed balance is preserved while reducing the long autocorrelation associated with local Metropolis moves. For spherical geometry, instead of the usual stereographic coordinate $z=\exp({i\phi})\tan(\theta/2)$, which sends the south pole to infinity, we employ the so-called double stereographic projection, which includes an extra inversion when the electron is in the northern hemisphere,
\begin{equation}
z(\theta,\phi)\rightarrow
\begin{cases}
z=\exp(i\phi)\tan(\theta/2), & \theta\le \pi/2,\\
w=\exp(i\phi)\cot(\theta/2)=1/z^{*}, & \theta>\pi/2,
\end{cases}
\end{equation}
thereby mapping the northern and southern hemispheres to two compact unit disks. In the spherical Laughlin wavefunction, this compactification removes points at infinity, producing smoother HMC forces and more stable, efficient sampling on the sphere.

\section{Numerical computation of the braiding matrix}
\label{sec:braiding_matrix}
We consider the general case where the positions of the quasiholes are collectively denoted by
$\mathbf{R}$.  The corresponding Moore--Read quasihole states span a
two-dimensional fusion space.  A braiding process is represented by an
adiabatic trajectory
\begin{equation}
    \mathbf{R}=\mathbf{R}(t),
    \qquad
    t\in[t_{0},t_{N_s}],
\end{equation}
which is closed in the configuration space of identical quasiholes.  The
trajectory is discretized into $N_s$ steps,
\begin{equation}
    \mathbf{R}_{n}\equiv\mathbf{R}(t_n),
    \qquad
    t_0<t_1<\cdots<t_{N_s}.
\end{equation}

When the braid permutes identical quasiholes, the final configuration is
identified with the initial one using the same quasihole-labeling convention
employed in defining the fusion-space basis.

At every point along the trajectory, we introduce a continuous orthonormal
basis
\begin{equation}
    \left\{
    \lvert\psi_{1}(t)\rangle,
    \lvert\psi_{2}(t)\rangle
    \right\}.
\end{equation}

As illustrated in Ref.~\cite{Tserkovnyak2003,wang2026hybrid}, for two neighboring configurations, we evaluate the overlap matrix
\begin{equation}
    M^{(n)}_{ab}
    =
    \langle\psi_{a}(t_n)
    \vert\psi_{b}(t_{n+1})\rangle,
    \qquad a,b=1,2,
    \label{eq:neighboring_overlap}
\end{equation}
and construct its anti-Hermitian part,
\begin{equation}
    A(t_n)
    =
    M^{(n)}-\left[M^{(n)}\right]^{\dagger},
    \qquad
    A(t_n)_{ab}
    =
    \langle\psi_a(t_n)\vert\psi_b(t_{n+1})\rangle
    -\mathrm{h.c.}
    \label{eq:discrete_connection}
\end{equation}
The overlaps in Eq.~\eqref{eq:neighboring_overlap} may be evaluated by Monte
Carlo sampling of the many-body wavefunctions.  The evolution operator is
then accumulated as,
\begin{align}
    U(t_{n+1})
    &=
    U(t_n)\,C_n,
    \label{eq:braid_recursion}
    \\
    C_n
    &=
    \left(
        \mathbbm{1}
        +\frac{1}{2}A(t_n)
    \right)
    \left(
        \mathbbm{1}
        -\frac{1}{2}A(t_n)
    \right)^{-1},
    \label{eq:cayley_step}
    \\
    U(t_0)
    &=
    \mathbbm{1}.
\end{align}
Since $A(t_n)^{\dagger}=-A(t_n)$, each $C_n$ is unitary.  Consequently,
unitarity is preserved at every discretized step, rather than only in the
continuum limit.  The braiding matrix associated with the complete contour
$\mathcal{C}$ is
\begin{equation}
    U_{\mathcal{C}}
    =
    C_0 C_1\cdots C_{N_s-1}.
    \label{eq:general_braiding_matrix}
\end{equation}

A general $2\times2$ unitary matrix obtained in this way can be parameterized
as
\begin{equation}
    U_{\mathcal{C}}
    =
    e^{i\chi}
    \begin{pmatrix}
        e^{i\eta}\cos\left(\dfrac{\beta}{2}\right)
        &
        i e^{-i\epsilon/2}
        \sin\left(\dfrac{\beta}{2}\right)
        \\[4pt]
        i e^{i\epsilon/2}
        \sin\left(\dfrac{\beta}{2}\right)
        &
        e^{-i\eta}\cos\left(\dfrac{\beta}{2}\right)
    \end{pmatrix}.
    \label{eq:braiding_parameterization}
\end{equation}
Here $\chi$ is an overall phase, $\eta$ gives half of the relative phase
between the diagonal matrix elements, $\beta$ controls the magnitude of
fusion-channel mixing, and $\epsilon$ determines the relative phase of the
off-diagonal elements.

The matrix $U_{\mathcal{C}}$ depends on the chosen fusion-space basis.
For a periodic basis transformation $V(t_{N_s})=V(t_0)$, it transforms by
unitary conjugation,
\begin{equation}
    U_{\mathcal{C}}
    \longrightarrow
    V^{\dagger}(t_0)\,
    U_{\mathcal{C}}\,
    V(t_0).
\end{equation}

Moreover, the numerical result can contain an overall phase arising from the
Aharonov--Bohm contribution and from the parallel transport of the quasihole
spin on the sphere.  Both effects leave the ratio of the two eigenvalues
unchanged.  Denoting the eigenvalues by $\lambda_{1}$ and $\lambda_{2}$, with
their ordering fixed continuously, we therefore define
\begin{equation}
    e^{i\alpha}
    =
    \frac{\lambda_{1}}{\lambda_{2}}.
    \label{eq:eigenvalue_ratio}
\end{equation}
The angle $\alpha$ is independent of the overall phase and of unitary changes
of basis, and is therefore the most direct quantity for comparing the
numerical braid with its topological prediction.

A substantial simplification occurs when the entire quasihole configuration
is generated by a rigid rotation of the sphere about the $z$ axis.  Let the
total rotation angle be $\Phi$ and choose equally spaced increments
\begin{equation}
    \Delta\phi=\frac{\Phi}{N_s},
    \qquad
    \mathbf{R}_{n+1}
    =
    \mathcal{R}_{z}(\Delta\phi)\mathbf{R}_{n}.
\end{equation}
The rotational symmetry makes all
neighboring overlap problems equivalent.  Thus,
\begin{equation}
    A(t_n)=A_{\mathrm{rot}},
    \qquad
    C_n=C_{\mathrm{rot}},
\end{equation}
for every $n$, where
\begin{equation}
    C_{\mathrm{rot}}
    =
    \left(
        \mathbbm{1}
        +\frac{1}{2}A_{\mathrm{rot}}
    \right)
    \left(
        \mathbbm{1}
        -\frac{1}{2}A_{\mathrm{rot}}
    \right)^{-1}.
\end{equation}
The full braiding matrix is consequently obtained from a single
representative overlap calculation:
\begin{equation}
    U_{\mathrm{rot}}
    =
    \left[
        \left(
            \mathbbm{1}
            +\frac{1}{2}A_{\mathrm{rot}}
        \right)
        \left(
            \mathbbm{1}
            -\frac{1}{2}A_{\mathrm{rot}}
        \right)^{-1}
    \right]^{N_s}.
    \label{eq:rotational_braiding_matrix}
\end{equation}
The rotational construction therefore replaces $N_s$ independent
many-body-overlap evaluations by one overlap evaluation followed by a matrix
power.
\section{Finite-size analysis of the braiding matrix}
\label{sec:fitting}

\subsection{Angles compared with the CFT}

We characterize the numerically computed braiding matrix $U_{\mathcal{C}}$ by the angles defined in \Cref{eq:braiding_parameterization}, evaluated in the basis $\{\ket{1},\ket{\psi}\}$ labeled by the fusion channel of a chosen pair of quasiholes. Only two of the four angles are suitable for a comparison with the CFT. The overall phase $\chi$ contains the Aharonov-Bohm phase and the parallel transport of the quasihole spin, which depend on the system size and on the enclosed solid angle, and is therefore discarded. The angle $\epsilon$ depends on the phase convention of the basis states: a rephasing $V=\mathrm{diag}(e^{i\varphi_1},e^{i\varphi_\psi})$, under which $U_{\mathcal{C}}\to V^\dagger U_{\mathcal{C}}V$, leaves $\chi$, $\eta$, and $\beta$ unchanged and shifts $\epsilon\to\epsilon+2(\varphi_1-\varphi_\psi)$. The angles $\eta$ and $\beta$, by contrast, are fixed once the fusion-channel basis is specified, and are obtained directly from the matrix elements,
\begin{equation}
    \eta=\frac{1}{2}\arg\frac{U_{11}}{U_{22}},
    \qquad
    \beta=2\arctan\sqrt{\frac{\left|U_{12}U_{21}\right|}{\left|U_{11}U_{22}\right|}}.
    \label{eq:eta_beta_extraction}
\end{equation}
For a unitary matrix, \Cref{eq:eta_beta_extraction} reproduces \Cref{eq:braiding_parameterization}; the symmetric form of $\beta$ treats both diagonal and both off-diagonal elements on the same footing and is therefore insensitive to small deviations from unitarity within the statistical errors. The Monte Carlo (MC) errors of the matrix elements are propagated to $\eta$ and $\beta$.

Unlike $\alpha$, the angles $\eta$ and $\beta$ are not invariant under a general unitary change of basis. For instance, the exchange in the classical scheme is described by $B^{(12)}$ or by $B^{(23)}$, depending on which branches of the fusion tree the exchanged quasiholes are identified with; the two matrices are related by $F^{\sigma}_{\sigma\sigma\sigma}$ and have $\beta=0$ and $\beta=\pi/2$, respectively. A meaningful comparison therefore requires the numerical basis to coincide with the fusion-channel basis in which the CFT matrices are written. This is confirmed by the modulus of the diagonal element, $|U_{11}|\approx1$ for the classical scheme and $|U_{11}|\approx1/\sqrt{2}$ for the rotational scheme over the entire range of system sizes, as expected from \Cref{eq:classical,eq:rotational 1}. Writing \Cref{eq:classical} in the form of \Cref{eq:braiding_parameterization} gives $\chi=\pi/8$, $\eta=-\pi/4$, and $\beta=0$, with $\epsilon$ undefined since the off-diagonal elements vanish. For the rotational scheme, \Cref{eq:rotational 1} gives $\chi=-\pi/2$, $\eta=\pi/4$, $\beta=\pi/2$, and $\epsilon=-\pi/2$, while the alternative representation in \Cref{eq:rotational 2} gives the same $\chi$, $\eta$, and $\beta$ with $\epsilon=\pi/2$. The comparison of $\eta$ and $\beta$ is therefore unaffected by the choice between the two representations in Fig.~S2.

The eigenvalues of \Cref{eq:braiding_parameterization} are $\lambda_\pm=e^{i\chi}e^{\pm i\theta}$ with $\cos\theta=\cos(\beta/2)\cos\eta$, independent of $\epsilon$. Hence, for $0\le\beta\le\pi$ and $|\eta|\le\pi/2$, the eigenvalue phase $e^{i\alpha}=\lambda_1/\lambda_2$ satisfies
\begin{equation}
    \cos\frac{\alpha}{2}=\cos\frac{\beta}{2}\cos\eta,
    \label{eq:alpha_eta_beta}
\end{equation}
which gives $\alpha=-\pi/2$ and $\alpha=-2\pi/3$ for the two sets of CFT values above, with the sign fixed by the ordering of the eigenvalues. \Cref{eq:alpha_eta_beta} also determines how finite-size errors of $\eta$ and $\beta$ propagate to $\alpha$. Near $\beta=0$ (classical scheme), $|\alpha|=2|\eta|+\mathcal{O}(\beta^2)$, so that $\alpha$ follows $\eta$ and is insensitive to a small mixing. Near $(\eta,\beta)=(\pi/4,\pi/2)$ (rotational scheme), $\delta|\alpha|=(\delta\beta+2\delta\eta)/\sqrt{3}$, so that both angles contribute at linear order.

\subsection{Distance dependence and fitting procedure}

Finite-size corrections are analysed as functions of the chord distance between neighboring quasiholes,
\begin{equation}
    d=2R\sin\frac{\Theta}{2},
    \qquad
    R=\sqrt{S}\,\ell_B=\sqrt{N-\tfrac{1}{2}}\,\ell_B,
    \label{eq:chord_distance}
\end{equation}
where $2S=2N-1$ for four quasiholes and $\Theta$ is the angle subtended by two neighboring quasiholes at the center of the sphere: $\Theta=\pi/2$ for the square of the classical scheme and $\Theta=\arccos(-1/3)=\pi/2+\arcsin(1/3)$ for the tetrahedron of the rotational scheme. For $N=10,20,\dots,60,80,100$, the data cover $4.36\,\ell_B\le d\le14.11\,\ell_B$ (classical) and $5.03\,\ell_B\le d\le16.29\,\ell_B$ (rotational).

Both angles are fitted to an exponentially damped oscillation,
\begin{equation}
    f(d)=f_\infty+A\,e^{-(d-d_0)/\xi}\cos\left(kd-\vartheta\right),
    \label{eq:fit_form}
\end{equation}
where $f$ stands for $\eta$ or $\beta$, $f_\infty$ is the asymptotic value, $\xi$ the decay length, $k$ the wave number, and $\vartheta$ the phase of the oscillation. The amplitude $A$ is defined at the smallest separation $d_0$ of each scheme, $d_0=4.359\,\ell_B$ (classical) and $5.033\,\ell_B$ (rotational), where it is directly constrained by the data, rather than extrapolated to $d=0$; this substantially reduces its correlation with $\xi$. We impose $A\ge0$, $\xi>0$, $0\le\vartheta<2\pi$, and $k\le\pi/\overline{\Delta d}$, where $\overline{\Delta d}$ is the mean spacing of the data in $d$, which excludes aliased solutions with more than half an oscillation period between adjacent data points. \Cref{eq:fit_form} is phenomenological and does not describe the data to the level of the MC errors; this is accounted for by the weighting described below. All quoted uncertainties are one standard deviation.

We perform two fits. In the joint fit, all four data sets ($\eta$ and $\beta$ for both schemes) share a single $\xi$ and a single $k$, while $A$, $\vartheta$, and $f_\infty$ are independent for each data set, giving $14$ parameters for $32$ data points. Each data set is weighted by the root-mean-square residual of a separate fit of that data set alone, which accounts for the deviations from \Cref{eq:fit_form} that exceed the MC errors. The resulting curves are shown in Fig.~1(c) of the main text. As a cross-check, we also fit each scheme separately, with $\xi$ and $k$ shared only between $\eta$ and $\beta$ of the same scheme. In this case, the propagated MC errors are used, rescaled by $\sqrt{\chi^2/\mathrm{dof}}$ of a separate fit of each angle. The parameters of both fits are listed in \Cref{tab:fit}.

\begin{table}[htb]
\caption{Parameters of \Cref{eq:fit_form} for the angles $\eta$ and $\beta$. Upper block: joint fit of all four data sets with common $\xi$ and $k$, shown in Fig.~1(c) of the main text. Lower blocks: separate fits of each scheme, with $\xi$ and $k$ shared between $\eta$ and $\beta$ of the same scheme. $A$ is the amplitude at the smallest separation $d_0$. All angles are in radians.}
\label{tab:fit}
\begin{ruledtabular}
\begin{tabular}{llcccc}
Scheme & Angle & $A$ & $\vartheta$ & $f_\infty$ & CFT \\
\hline
\multicolumn{6}{l}{Joint fit: $\xi=2.768\pm0.059\,\ell_B$, $k=0.964\pm0.016\,\ell_B^{-1}$, $\chi^2/\mathrm{dof}=40/18$} \\
Classical & $\eta$ & $0.3616\pm0.0036$ & $1.05\pm0.11$ & $-0.7858\pm0.0015$ & $-\pi/4\approx-0.7854$ \\
 & $\beta$ & $0.112\pm0.026$ & $4.23\pm0.45$ & $0.069\pm0.010$ & $0$ \\
Rotational & $\eta$ & $0.044\pm0.012$ & $4.27\pm0.43$ & $0.7928\pm0.0032$ & $\pi/4\approx0.7854$ \\
 & $\beta$ & $0.144\pm0.036$ & $2.71\pm0.19$ & $1.5768\pm0.0062$ & $\pi/2\approx1.5708$ \\
\hline
\multicolumn{6}{l}{Separate fit: $\xi=2.771\pm0.070\,\ell_B$, $k=0.983\pm0.018\,\ell_B^{-1}$, $\chi^2/\mathrm{dof}=6.2/8$} \\
Classical & $\eta$ & $0.3625\pm0.0090$ & $1.18\pm0.12$ & $-0.7841\pm0.0015$ & $-\pi/4\approx-0.7854$ \\
 & $\beta$ & $0.110\pm0.022$ & $4.34\pm0.31$ & $0.073\pm0.007$ & $0$ \\
\hline
\multicolumn{6}{l}{Separate fit: $\xi=2.8\pm1.0\,\ell_B$, $k=1.85\pm0.13\,\ell_B^{-1}$, $\chi^2/\mathrm{dof}=7.7/8$} \\
Rotational & $\eta$ & $0.043\pm0.017$ & $3.56\pm0.99$ & $0.7924\pm0.0036$ & $\pi/4\approx0.7854$ \\
 & $\beta$ & $0.20\pm0.12$ & $4.98\pm0.83$ & $1.5788\pm0.0080$ & $\pi/2\approx1.5708$ \\
\end{tabular}
\end{ruledtabular}
\end{table}

\subsection{Results}

Both fits reproduce the CFT values. In the joint fit, the classical scheme gives $\eta_\infty=-0.7858\pm0.0015$, in agreement with $-\pi/4$ within one standard deviation. For the rotational scheme, $\eta_\infty=0.7928\pm0.0032$ and $\beta_\infty=1.5768\pm0.0062$ deviate from $\pi/4$ and $\pi/2$ by $0.9\%$ and $0.4\%$, corresponding to $2.3$ and $1.0$ standard deviations. These deviations are comparable to the point-to-point scatter of the rotational data at large $d$, $\approx0.01$, which \Cref{eq:fit_form} does not capture. Through \Cref{eq:alpha_eta_beta}, the extrapolated eigenvalue phases are $\alpha_\infty=-1.573\pm0.003$ (classical) and $\alpha_\infty=-2.106\pm0.005$ (rotational), compared with $-\pi/2\approx-1.571$ and $-2\pi/3\approx-2.094$, where correlations between the fitted asymptotes are neglected. The asymptotic values of the joint and separate fits agree within their uncertainties.

The classical scheme yields a small but nonzero $\beta_\infty=0.069\pm0.010$, corresponding to an off-diagonal weight $\sin^2(\beta_\infty/2)\approx1.2\times10^{-3}$. Since $\beta$ is non-negative by construction, any residual off-diagonal amplitude, e.g., from the incomplete orthogonality of the finite-size conformal blocks, appears as a positive offset rather than averaging to zero. The largest system ($d\approx14.1\,\ell_B$) gives $\beta\approx0.02$, well below the fitted asymptote, so that the nonzero $\beta_\infty$ more likely reflects the limited flexibility of \Cref{eq:fit_form} than a genuine asymptotic mixing. In any case, by \Cref{eq:alpha_eta_beta} its effect on $\alpha$ is of second order, $|\delta\alpha|\approx1\times10^{-3}$.

The decay length is common to both schemes and both angles: the separate fits give $\xi=2.771\pm0.070\,\ell_B$ (classical) and $\xi=2.8\pm1.0\,\ell_B$ (rotational), in agreement with the joint fit, $\xi=2.768\pm0.059\,\ell_B$. The wave numbers are less consistent, $k=0.983\pm0.018\,\ell_B^{-1}$ (classical) and $k=1.85\pm0.13\,\ell_B^{-1}$ (rotational). The latter is poorly determined: the oscillation amplitudes of the rotational scheme are small, the fitted $k$ lies close to its upper bound $\pi/\overline{\Delta d}\approx1.95\,\ell_B^{-1}$, and relaxing the fit bounds yields $k\approx1.2$--$1.3\,\ell_B^{-1}$ with a comparable $\chi^2$. Accordingly, the larger $\chi^2/\mathrm{dof}\approx2.2$ of the joint fit indicates that a common $k$ is only an approximate description.

Because the two schemes share the same $\xi$, the ratio of their oscillation envelopes is independent of $d$. For $\eta$, the classical envelope exceeds the rotational one by a factor $(0.362/0.044)\,e^{-(5.033-4.359)/\xi}\approx6.5$ at equal separation. For $\beta$, the envelopes are comparable, with the rotational one, if anything, slightly larger (ratio $0.6\pm0.2$). The improved robustness of the rotational scheme therefore resides primarily in the relative phase $\eta$ between the two fusion channels. Since $\beta\approx0$ in the classical scheme, its eigenvalue phase follows $|\alpha|\approx2|\eta|$ and inherits deviations of $\approx0.7$ at the smallest separation. In the rotational scheme, deviations of $\eta$ and $\beta$ enter through $\delta|\alpha|=(\delta\beta+2\delta\eta)/\sqrt{3}$, and $\alpha$ remains within $\approx0.05$ of $-2\pi/3$ for all system sizes.

\begin{figure}
    \centering
    \includegraphics[width=\linewidth]{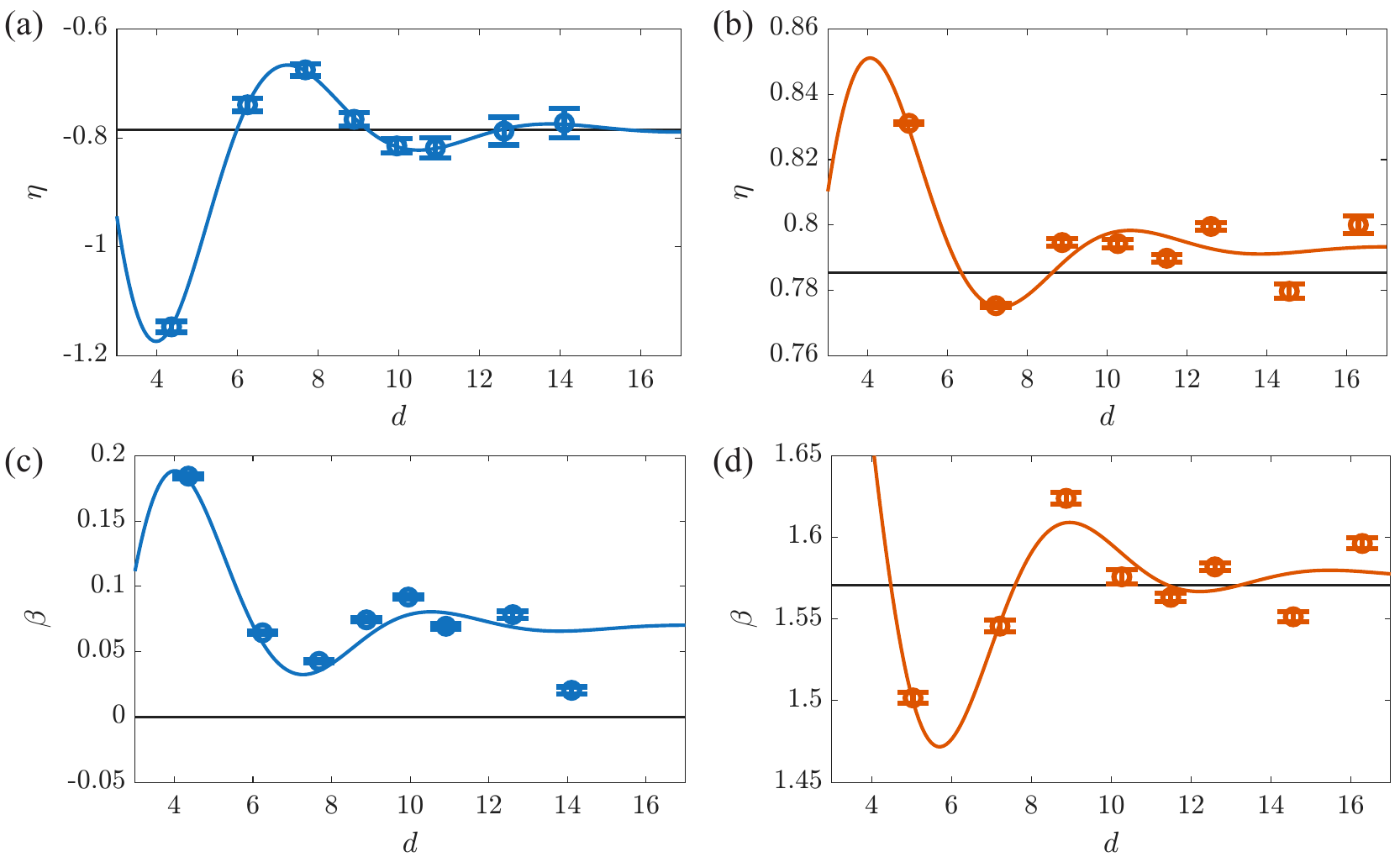}
    \caption{Finite-separation dependence of the angles $\eta$ [(a), (b)] and $\beta$ [(c), (d)] of \Cref{eq:braiding_parameterization} for the classical (left, blue) and rotational (right, orange) schemes, as functions of the chord distance $d$ of \Cref{eq:chord_distance}. Error bars are MC errors. The curves are the joint fit of \Cref{eq:fit_form} with common $\xi$ and $k$ (\Cref{tab:fit}), and the horizontal lines are the CFT values.}
    \label{fig:scheme_compare_zoom}
\end{figure}

\bibliographystyle{longapsrev4-2}
\bibliography{ref}